\documentclass[final,1p,times]{elsarticle}

\usepackage{graphicx}
\usepackage{amssymb}
\usepackage{enumerate}
\usepackage{multirow}
\usepackage{lscape}
\usepackage{setspace}
\usepackage{array}
\usepackage{booktabs}
\usepackage{hhline}
\usepackage{color}
\usepackage{colortbl}

\newcolumntype{C}[1]{>{\centering\let\newline\\\arraybackslash\hspace{0pt}}p{#1}}
\newcolumntype{P}[1]{>{\raggedright\arraybackslash}p{#1}}

\definecolor{Gray}{gray}{0.7}

\newcommand{\f}{\frac}
\newcommand{\Msun}{\,\ifmmode {\rm M}_\odot \else ${\rm M}_\odot$\fi}

\newcommand{\cg}{\cellcolor{Gray}}

\biboptions{sort&compress}

\journal{Physics of the Dark Universe}

\begin{document}

\begin{frontmatter}

\title{Numerical Simulations of the Dark Universe: \\State of the Art and the Next Decade}

\author[label1]{Michael Kuhlen}
\ead{mqk@astro.berkeley.edu}
\author[label2]{Mark Vogelsberger}
\ead{mvogelsb@cfa.harvard.edu}
\author[label3,label4]{Raul Angulo}
\ead{reangulo@slac.stanford.edu}

\address[label1]{Theoretical Astrophysics Center, University of California Berkeley, Hearst Field Annex, Berkeley, CA 94720, USA}
\address[label2]{Hubble Fellow, Harvard-Smithsonian Center for Astrophysics, 60 Garden Street, Cambridge, MA 02138, USA}
\address[label3]{Max-Planck-Institute for Astrophysics, Karl-Schwarzschild-Str. 1, 85740 Garching, Germany}
\address[label4]{Kavli Institute for Particle Astrophysics and Cosmology, Stanford University, Menlo Park, CA 94025, USA}

\begin{abstract}
We present a review of the current state of the art of cosmological dark matter simulations, with particular emphasis on the implications for dark matter detection efforts and studies of dark energy. This review is intended both for particle physicists, who may find the cosmological simulation literature opaque or confusing, and for astro-physicists, who may not be familiar with the role of simulations for observational and experimental probes of dark matter and dark energy.  Our work is complementary to the contribution by M.~Baldi in this issue, which focuses on the treatment of dark energy and cosmic acceleration in dedicated N-body simulations.

Truly massive dark matter-only simulations are being conducted on national supercomputing centers, employing from several billion to over half a trillion particles to simulate the formation and evolution of cosmologically representative volumes (cosmic scale) or to zoom in on individual halos (cluster and galactic scale). These simulations cost millions of core-hours, require tens to hundreds of terabytes of memory, and use up to petabytes of disk storage. Predictions from such simulations touch on almost every aspect of dark matter and dark energy studies, and we give a comprehensive overview of this connection. We also discuss the limitations of the cold and collisionless DM-only approach, and describe in some detail efforts to include different particle physics as well as baryonic physics in cosmological galaxy formation simulations, including a discussion of recent results highlighting how the distribution of dark matter in halos may be altered. We end with an outlook for the next decade, presenting our view of how the field can be expected to progress.

\end{abstract}


\end{frontmatter}

\section{Introduction}

For almost 80 years now \cite{zwicky_rotverschiebung_1933} astronomers have been accumulating evidence that the dominant form of matter in the universe is dark and non-baryonic. Just a bit more than a decade ago, we obtained solid observational evidence of cosmic acceleration \cite{riess_observational_1998,perlmutter_measurements_1999}, requiring yet another mysterious contribution to the total energy budget of the universe, dark energy. The conclusion is staggering: we live in a universe that is energetically dominated by dark matter (DM) and dark energy (DE). For DM at least we have a well-motivated theoretical framework, in which it is comprised of fundamental particles predicted to arise in extensions of the standard model of particle physics like supersymmetry \cite{bertone_particle_2005}. There are many ongoing and proposed efforts to obtain experimental confirmation of the hypothesis of particle DM, for example through so called indirect (annihilation/decay) and direct (nuclear scattering) detection. These signatures depend on the detailed distribution of DM throughout the universe, from cosmic all the way down to sub-galactic scales. For DE, on the other hand, we have only a very rudimentary theoretical understanding, and are a firmly in an exploratory phase, conducting and designing future surveys and measurements that will provide us with additional clues to its nature.

The last 40 years have seen tremendous progress in our understanding of cosmic structure and galaxy formation. Much of this understanding has come at the hand of beautifully simple analytic arguments and insights. The calculation of the Cold Dark Matter (CDM) power spectrum \cite{peebles_large-scale_1982,blumenthal_formation_1984}, Press \& Schechter theory \cite{press_formation_1974}, the statistics of peaks in Gaussian random fields \cite{bardeen_statistics_1986}, and White \& Rees' galaxy formation model \cite{white_core_1978} are a few seminal examples. Yet, it is clear that purely analytical approaches have arrived at the limits of their reach. Fueled by continuing advances in numerical methods and computational capabilities, the future of structure formation and galaxy formation theory is going to be led by numerical simulations.

Indeed, in the last few decades Moore's law combined with heavy infrastructure investments has already tremendously increased available computing resources, and the field of computational cosmology has taken full advantage. Full-box simulations of substantial fractions of the observable Universe are now being conducted with over half a trillion particles, and zoom-in simulations of individual halos have exceeded the billion particle level. In this review we present a snapshot of the current state of the art of cosmological numerical simulations, with a particular emphasis on the DM and DE problems. Its intended audience is both the particle physicist with an interest in DM and DE, who may find the simulation literature to be opaque and confusing, as well as the astro-physicist, who may not be up to speed with observational and experimental probes of DM and DE or with the importance of simulations for their interpretation. Our work is complementary to the contribution by M.~Baldi in this issue \cite{baldi_dark_2012}, which focuses on the treatment of DE and cosmic acceleration in dedicated N-body simulations.

This review is structured as follows. We begin in \S\ref{sec:DMsims} with a review of the domain of DM simulations. What sorts of predictions do they make? How are these predictions relevant for DM detection efforts and for probes of DE? We cover astrophysical probes, indirect, and direct detection of DM, as well as probes of DE in the form of baryon acoustic oscillations, redshift-space distortions, cluster mass functions, and weak gravitational lensing. In \S\ref{sec:state-of-the-art} we present a survey of the state of the art in late 2012 of cosmological DM simulations on cosmic, cluster, and galactic scales. We give an overview of the most commonly employed numerical techniques, then describe some of the largest simulations run to date, and the computational resources they used. We discuss the limits of the cold and collisionless DM-only approach, and efforts to go beyond it by simulating alternative DM physics. We include an discussion of hydrodynamic galaxy formation simulations, cover numerical methods, algorithmic and technical difficulties, and highlight some recent results regarding how the DM distribution in halos may be altered by baryonic physics. Finally, in \S\ref{sec:next_decade} we present our vision of this field for the next decade. What are the important questions to tackle, and how best to do so? What developments should be pursued in order to take advantage of technological advances?

\section{Dark Matter Simulations and the Dark Universe}\label{sec:DMsims}

\begin{figure}
\centering
\includegraphics[width=\textwidth]{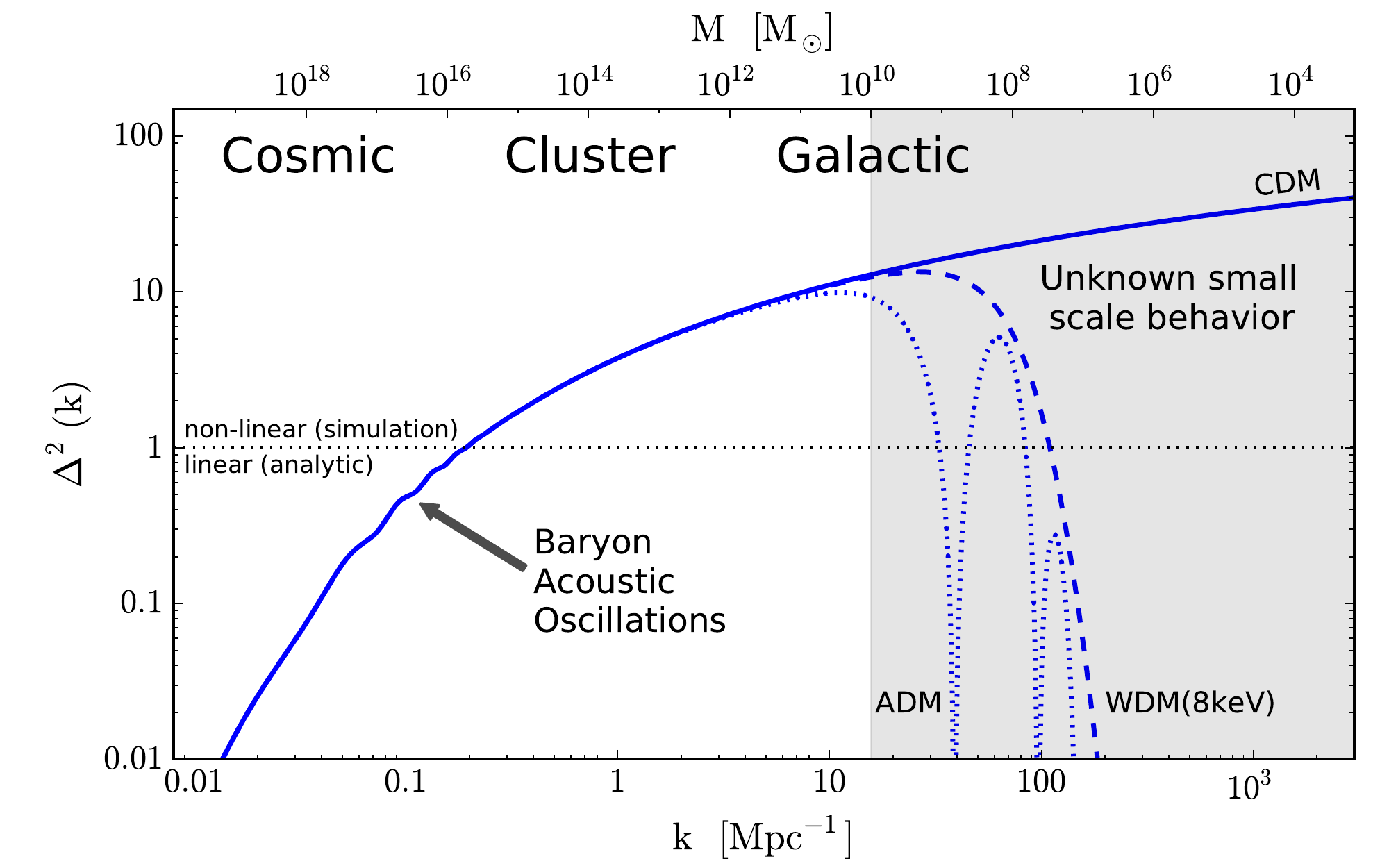}
\caption{$\Delta^2(k) \equiv 4\pi (k/2\pi)^3 P(k)$, the linear power spectrum of density fluctuations at $z=0$. The solid line is the canonical cold DM model with an Eisenstein \& Hu (1997) \cite{eisenstein_baryonic_1998} transfer function. The dashed line is a thermal relic warm DM model with $m_{\rm WDM} = 8$ keV \cite{bode_halo_2001}. The dotted line is an atomic DM model \cite{cyr-racine_cosmology_2012}. We used WMAP7 cosmological parameters \cite{jarosik_seven-year_2011}, $\Omega_m = 0.265$, $\Omega_\Lambda = 0.735$, $\Omega_b = 0.0449$, $h=0.71$, $\sigma_8 = 0.801$, and $n_s=0.963$.}
\label{fig:Delta2}
\end{figure}

The numerical simulation discussed in this review together span an enormous range of length scales, more than 8 orders of magnitude reaching from near horizon scale ($\sim$ 20 Gpc) down to sub-Galactic (tens of pc). Individually they focus on different regimes (see \S\ref{sec:state-of-the-art} and Table~\ref{tab:DM_simulations}), but all have in common that they evolve the growth of DM density fluctuations all the way to the present epoch at redshift zero.\footnote{We deliberately omit from our discussion multi-billion particle simulations that focus only on the first billion years of cosmic evolution, for studying the epoch of reionization \cite{iliev_can_2012} or early supermassive black hole growth \cite{degraf_growth_2012}.} 

The shape of the CDM power spectrum results in a hierarchical, bottom-up process of structure formation, in which small and low mass objects collapse first and over time merge to form ever more massive structures, until the onset at $z \approx 1$ of DE induced accelerated expansion begins to halt further collapse. In Fig.~\ref{fig:Delta2} we show a plot of the \textit{linear} dimensionless matter power spectrum $\Delta^2(k) \equiv 4\pi (k/2\pi)^3 P(k)$ at $z=0$ versus the wavenumber $k$ of the fluctuation. Where $\Delta \gtrsim 1$, gravitational collapse will have proceeded to the non-linear regime and typical objects of the corresponding mass will have collapsed. Cosmic scales, including the Baryon Acoustic Oscillation feature discussed in \S\ref{sec:relevance_de}.i, remain in the linear or mildly non-linear regime, while cluster and galactic scales are strongly non-linear. Note that computational demands grow strongly with the degree of non-linearity resolved in the simulation.

Observational constraints from the Cosmic Microwave Background, cluster abundances, galaxy clustering, weak lensing and the Lyman-$\alpha$ forest have constrained the power spectrum of density fluctuations and provide a remarkably good agreement with the predictions of $\Lambda$CDM cosmology. On smaller scales (shaded gray in Fig.~\ref{fig:Delta2}) we currently do not have robust observational constraints, and here numerical simulations typically rely on extrapolations under the assumption of CDM theory. As we discuss in \S\ref{sec:departures}, different assumptions are plausible and are the subject of many ongoing investigations. As an example, we show two alternative models with a suppression of small scale power, a warm DM \cite{bode_halo_2001} and an atomic DM \cite{cyr-racine_cosmology_2012} model.

\subsection{Domain}

In the following we provide a brief summary of the domain of cosmological DM simulations, roughly organized from large scales to small.\footnote{Also see \cite{diemand_structure_2011} for a recent review of simulation results concerning the evolution and structure of CDM halos.} This is not meant to be an exhaustive review of all current results, but rather an overview of those results with particular relevance for DM and DE experiments. The references that we list provide a jumping-off point for further reading.

\begin{enumerate}

  \item \underline{Large Scale Structure}
    
    The largest scale density fluctuations in the universe never grow beyond
mildly non-linear, and even early CDM simulations predicted
that the large scale distribution of DM in the universe is not completely
homogeneous, instead exhibiting \textbf{voids, walls, and filaments} whose
statistical description is in remarkable agreement with the large scale
distribution of galaxies \cite[e.g.][]{springel_large-scale_2006}.

    Simulations spanning an ever larger fraction of the volume of the observable Universe at
increasingly high resolution have been able to quantify the DM density and velocity fields as
well as the \textbf{halo
mass function} together with the full hierarchy of \textbf{halo correlation functions}, and their
evolution with cosmic time
\cite{sheth_large-scale_1999,jenkins_mass_2001,warren_precision_2006,tinker_toward_2008,reed_toward_2012}.

  \item \underline{Individual isolated halo properties}

    On the scale of individual halos, DM-only numerical simulations have measured \textbf{halo shapes} to show significant departures from sphericality, with halos typically being prolate and increasingly so towards their centers. Major-to-minor axis ratios of 2 or greater are not uncommon, and more massive halos tend to be less spherical than lower mass halos \cite{allgood_shape_2006,bett_spin_2007}. Shapes and {\bf kinematics} seem to be closely connected. While the spherically averaged anisotropy profile ($\beta(r) = 1 - 0.5 \, \sigma_t^2 / \sigma_r^2$) grows from zero (isotropic) to about 0.4 (mild radial anisotropy) \cite{ghigna_density_2000,diemand_velocity_2004}, the local $\beta$ values correlate with halo shape: positive (radial) on the major axis and negative (tangential) on the minor axis \cite{zemp_graininess_2009,sparre_asymmetric_2012}.

    The DM mass distribution within halos is well described by a near-universal \textbf{density profile}, the so-called NFW profile \cite{navarro_universal_1997}, which has the form of a double-power-law with the logarithmic slope $\gamma \equiv {\rm d}\!\log\rho/{\rm d}\!\log r$ transitioning at the scale radius $r_s$ from $\gamma = -3$ at large radii to $\gamma = -1$ in the center. More recent higher resolution simulations, however, have found a central slope shallower than $\gamma = -1$, indicating that the density profile may be better described by a functional form with a central slope gradually flattening to $\gamma = 0$, e.g. the Einasto profile \cite{navarro_diversity_2010,stadel_quantifying_2009}.

    The scaling of the transition radius $r_s$ with halo mass, formation time, and environment is typically described in terms of a ``concentration'', defined as the ratio of the virial radius to the scale radius, $c=R_{\rm vir} / r_s$. DM simulations have quantified the \textbf{concentration-mass relationship}, its scatter, and its evolution with time \cite{bullock_profiles_2001,wechsler_concentrations_2002,zhao_accurate_2009}. Concentrations typically increase for lower mass halos, presumably reflecting their earlier collapse times when the mean density of the universe was higher, although recent work has reported an upturn of concentrations at high masses \cite{prada_halo_2012} presumably caused by out-of-equilibrium systems \cite{ludlow_dynamical_2012}.

    Lastly, we mention the remarkable finding from simulations that the
\textbf{pseudo-phase-space profile}, the ratio of the spherically averaged DM mass density to the
cube of its spherically averaged radial velocity dispersion, is well described by a single power-law, $Q(r)
\equiv \rho(r)/\sigma_{r}(r)^3 \sim r^{-1.84}$, even though neither the density nor
the velocity dispersion profiles by themselves are
\cite{taylor_phase-space_2001,dehnen_dynamical_2005,schmidt_alas_2008,stadel_quantifying_2009}.
The power law slope is remarkably close to analytic predictions based on
spherical secondary-infall similarity solution
\cite{bertschinger_self-similar_1985} and their generalization
\cite{vogelsberger_non-spherical_2011} in the inner, virialized regions of halos
\cite{ludlow_secondary_2010}. Departures from a pure power-law occur around the
virial radius, close to the location of first shell crossing, where
particles have not yet fully virialized. Note also that the low velocity dispersion in subhalos leads to large fluctuations in local estimates of the phase-space density and thus its spherical average does not follow a single power law \cite{stadel_quantifying_2009,ludlow_secondary_2010}.

  \item \underline{Substructure}

    The numerical resolution achievable by state of the art DM-only simulations has grown to the point where it has become possible to follow bound DM structures beyond their merging with a larger halo. This has allowed studies of DM substructure, consisting both of a population of surviving self-bound subhalos orbiting within the potential of their hosts and the debris associated with their tidal stripping and disruption.

    These simulations have been able to probe the \textbf{subhalo mass function} and $\mathbf{V_{\rm max}}$ \textbf{function} over $\sim 5$ of magnitude in subhalo mass \cite{diemand_clumps_2008,springel_aquarius_2008}, and have shown that they are well fit by simple power laws, ${\rm d}N/{\rm d}M \sim M^{-1.9}$ and $N(>V_{\rm max}) \sim V_{\rm max}^{-3}$. It has even been possible to resolve up to 4 levels of the \textbf{sub-substructure hierarchy} \cite{springel_aquarius_2008}, but the statistics are currently not sufficient to quantify sub-substructure scaling laws.

    As with isolated halos, determining \textbf{subhalo density profiles and concentrations} is of great interest \cite{diemand_clumps_2008,springel_aquarius_2008}. At current resolutions, subhalo density profiles appear to be well fit by both NFW or Einasto profiles, and there is some evidence for a radial scaling of subhalo concentration, with higher concentrations for subhalos closer to the center \cite{diemand_formation_2007,springel_aquarius_2008}. The latter effect is likely due to a combination of stronger tidal forces and the earlier collapse times of subhalos found close to the center \cite{diemand_distribution_2005}. The \textbf{spatial distribution} of subhalos within their hosts appears to be ``anti-biased'' with respect to the host's mass distribution \cite{ghigna_density_2000,diemand_velocity_2004,diemand_formation_2007,kuhlen_shapes_2007,springel_aquarius_2008, angulo_fate_2009}, meaning that the subhalo density normalized by the host's mass density profile decreases towards the center. The degree of this anti-bias depends on how the subhalo sample is selected: a current mass-selected sample is more affected by tidal stripping, which is stronger closer to the host's center, resulting in a pronounced anti-bias than a for sample that is selected by properties unaffected by tidal stripping, like the mass before accretion \cite{nagai_radial_2005,faltenbacher_velocity_2006}.

    \textbf{Velocity-space substructure}, in the form of tidal streams and debris flow, is another topic that has received attention \cite{vogelsberger_phase-space_2009,zemp_graininess_2009,kuhlen_dark_2010,kuhlen_direct_2012}. DM tidal streams have low configuration space density, typically only a few per cent of the underlying host halo density. However, since they are significantly colder than the host halo, they have a high phase-space density contrast. Another form of velocity-space substructure may be contributed by a so-called \textbf{dark disk} \cite{read_dark_2009}, a component of the DM halo that is co-rotating with the stellar disk, which may have been deposited by satellites disrupted in the plane of the Galaxy.

   On smallest scales the \textbf{fine-grained phase-space structure} describes the detailed distribution of DM in configuration and velocity space. Before the onset of nonlinear structure formation CDM was almost uniformly distributed with particles lying on a thin three-dimensional hypersurface embedded in six-dimensional phase-space. Due to their collisionless character DM particles then follow the Vlasov-Poisson equations leading to stretching and folding of this initial sheet. Therefore, at later times the velocity distribution of DM at a given point in configuration space is a superposition of fine-grained streams of different velocities. Furthermore, phase-space fold catastrophes lead to caustics, where the DM configuration space density can become very large, many order of magnitudes larger than the mean halo density \cite{white_dark_2009,vogelsberger_streams_2011}. It has only very recently become possible to study these effects numerically by extending classical N-body schemes \cite{vogelsberger_fine-grained_2008, vogelsberger_caustics_2009}.

  \item \underline{Local DM}

    Lastly, numerical simulations provide expectations regarding the DM distribution at the solar circle, $\sim 8$ kpc from the Galactic Center. Simulations indicate that the \textbf{local density} of DM is likely to be quite smooth and uniform \cite{vogelsberger_phase-space_2009,kamionkowski_galactic_2010}, since strong tidal forces disrupt most subhalos close to the center. The flipside of this coin is that the local neighborhood should be crossed by many DM \textbf{tidal streams} \cite{vogelsberger_phase-space_2009,kuhlen_dark_2010}, cumulatively referred to as \textbf{debris flow} \cite{kuhlen_direct_2012}.
   
\end{enumerate}

\subsection{Relevance for Dark Matter Detection}

\begin{table}
\singlespacing
\centering
\scriptsize
 
\renewcommand{\arraystretch}{1.2}

\begin{tabular}{c|p{1.35in}!{\vrule width 1.5pt}c|c!{\vrule width 1.5pt}c|c|c|c!{\vrule width 1.5pt}c|c|c|c|c|c!{\vrule width 1.5pt}c|c|c!{\vrule width 1.5pt}}

\multicolumn{2}{c!{\vrule width 1.5pt}}{} & \multicolumn{2}{c!{\vrule width 1.5pt}}{\normalsize LSS} & \multicolumn{4}{c!{\vrule width 1.5pt}}{\normalsize Halos} & \multicolumn{6}{c!{\vrule width 1.5pt}}{\normalsize Substructure} & \multicolumn{3}{c!{\vrule width 1.5pt}}{\normalsize Local} \\ \cline{3-17}

\multicolumn{2}{c!{\vrule width 1.5pt}}{} & \rotatebox{90}{voids, walls, filaments} & \rotatebox{90}{halo mass functions} & \rotatebox{90}{concentration-mass relation~~} & \rotatebox{90}{halo shapes} & \rotatebox{90}{density profiles} & \rotatebox{90}{pseudo-phase-space density} & \rotatebox{90}{mass (or V$_{\rm max}$) functions} & \rotatebox{90}{density profiles} & \rotatebox{90}{central density} & \rotatebox{90}{spatial distribution} & \rotatebox{90}{streams} & \rotatebox{90}{folds \& caustics} & \rotatebox{90}{local density} & \rotatebox{90}{tidal streams} & \rotatebox{90}{dark disk} \\ \midrule[1.5pt]

\multirow{4}{*}{\rotatebox{90}{\scriptsize Astrophysical}} & Dwarf galaxy abundance & & & & \cg & \cg & & \cg & \cg & \cg & \cg & & & & \cg & \cg \\ \hhline{~*{16}{-}}
 & Dwarf galaxy kinematics & & & & & & & \cg & \cg & \cg & & & & & & \\ \hhline{~*{16}{-}}
 & Stellar streams & & & & \cg & \cg & & \cg & & & \cg & \cg & & & & \\ \hhline{~*{16}{-}}
 & Gravitational lensing & \cg & \cg & \cg & \cg & \cg & & \cg & \cg & & \cg & & & & & \\ \midrule[1.5pt]

\multirow{10}{*}{\rotatebox{90}{\normalsize Indirect Detection}} & Extra-galactic DGRB & \cg & \cg & \cg & & \cg & & \cg & \cg & & & & & & & \\ \hhline{~*{16}{-}}
 & Galactic DGRB & & & & \cg & \cg & & \cg & \cg & & \cg & & & & & \\ \hhline{~*{16}{-}}
 & Clusters & & & \cg & \cg & \cg & & \cg & \cg & & & & & & & \\ \hhline{~*{16}{-}}
 & Galactic Center & & & & & \cg & & & & & & & & & & \\ \hhline{~*{16}{-}}
 & Milky Way Dwarfs & & & \cg & & \cg & & & \cg & \cg & & & & & & \\ \hhline{~*{16}{-}}
 & Dark Subhalos & & & \cg & & \cg & & \cg & \cg & \cg & \cg & & & & & \\ \hhline{~*{16}{-}}
 & Local anti-matter & & & & & & & & & & & & & \cg & & \\ \hhline{~*{16}{-}}
 & Neutrinos from Earth \& Sun & & & & & & & & & & & & & \cg & \cg & \cg \\ \hhline{~*{16}{-}}
 & Substructure boost & & & & & & & \cg & \cg & & \cg & & \cg & & & \\  \hhline{~*{16}{-}}
 & Sommerfeld boost & & & & & & \cg & & & & & \cg & \cg & & \cg & \cg \\ \midrule[1.5pt]

\multirow{4}{*}{\rotatebox{90}{\normalsize Direct}} & ``Vanilla'' $\sim\!100$ GeV DM & & & & & & & & & & & & & \cg & & \\  \hhline{~*{16}{-}}
 & light / inelastic DM & & & & & & & & & & & & & \cg & \cg & \cg \\ \hhline{~*{16}{-}}
 & axions & & & & & & & & & & & & \cg & \cg & \cg & \cg \\ \hhline{~*{16}{-}}
 & directionally sensitive experiments & & & & & & & & & & & & & \cg & \cg & \cg  \\ \midrule[1.5pt]
\end{tabular}

\caption{A matrix showing which predictions from numerical DM simulations affect which astrophysical probes, indirect, and direct DM detection efforts, and vice versa.}
\label{tab:relevance}
\end{table}

Predictions from cosmological DM simulations affect nearly all DM detection efforts, in a variety of ways. In the following sections we highlight some of these inter-dependencies, which are also summarized in Table~\ref{tab:relevance}.

\subsubsection{Astrophysical Probes}\label{sec:astrophysical}

\begin{enumerate}
\item \textbf{Dwarf galaxies}

  The abundance of dwarf satellite galaxies orbiting our Milky Way and M31 is potentially sensitive to the nature of DM (see \S\ref{sec:CDM_challenges}). Results from CDM simulations have been used to predict how many more ultra-faint dwarf galaxies should be detected, once surveys more sensitive than the Sloan Digital Sky Survey (SDSS) and covering the southern hemisphere (e.g. DES, Skymapper, Pan-STARRS, LSST) come online \cite{tollerud_hundreds_2008}. The kinematics of stars in the very centers of dwarf spheroidal galaxies have been used to constrain the DM mass enclosed within their half-light radius ($\sim 300$ pc) \cite{strigari_common_2008,wolf_accurate_2010} and the shape of the DM density profile of their host halos, and these measurements have been confronted with the predictions from CDM simulations \cite{munoz_probing_2009,de_blok_core-cusp_2010,walker_method_2011,rashkov_assembly_2012}.

\item \textbf{Stellar streams}

  The discovery in the SDSS of extended stellar streams \cite{belokurov_field_2006}, arising from the tidal disruption of dwarf galaxies, has provided first-hand evidence for the hierarchical build-up of the Milky Way. DM counterparts to these stellar streams are seen in numerical simulations \cite{zemp_graininess_2009,cooper_galactic_2010}, and raise expectations that many more stellar streams remain to be discovered \cite{helmi_substructure_2011}. Additionally, cold stellar streams associated with the disruption of globular clusters \cite{grillmair_four_2009} are promising probes of the clumpiness of the Milky Way's DM halo, since interactions with passing subhalos should produce kinks and gaps the stream \cite{carlberg_star_2009,yoon_clumpy_2011}.

\item \textbf{Gravitational lensing}

  Gravitational lensing provides important probes of DM on cosmic, cluster, and galactic scales that can be compared to the predictions from numerical simulations. We can distinguish between weak lensing (see also \S\ref{sec:relevance_de}.iv), which can be used to tomographically map the large scale distribution of DM \cite{massey_dark_2007} and to measure the total masses and shapes of individual halos through cluster and galaxy-galaxy lensing \cite{mandelbaum_galaxy_2006,parker_masses_2007}, and strong lensing, which can probe the central slope of DM density profiles \cite{sand_dark_2004} and is sensitive, through flux ratio anomalies \cite{mao_evidence_1998} and potentially time-delay perturbations \cite{keeton_new_2009}, to the amount of DM substructure present in cluster and galaxy halos \cite{dalal_direct_2002,metcalf_compound_2001}. Recent studies comparing to predictions from numerical simulations tend to find that the amount of substructure present in DM halos may be insufficient to explain the observed occurence of flux ratio anomalies \cite{amara_simulations_2006,maccio_radial_2006,xu_effects_2009}. However, the effects of intervening line-of-sight structures can be important \cite{xu_effects_2012}.

\end{enumerate}

\subsubsection{Indirect Detection}\label{sec:indirect}

Indirect detection of DM refers to the search for the products of pair-annihilations of DM. The direct annihilation into two photons is typically loop-suppressed, and so the dominant annihilation channel is into quarks, gauge (or Higgs) bosons, or directly into leptons. The hadronization of heavy annihilation products results in gamma ray photons, electrons and positrons, and neutrinos. All of these are potentially observable, for example with ground based Atmospheric Cerenkov Telescopes (MAGIC, VERITAS, H.E.S.S.) and neutrino detectors (IceCube), balloon-borne detectors (ATIC), and space-based satellites (PAMELA, Fermi Gamma-ray Space Telescope) and experiments (AMS-02 on the International Space Station). In the following we discuss some of the possible DM annihilation signatures.

\begin{enumerate}
\item \textbf{Extra-galactic Diffuse Gamma-ray Background}

  The extra-galactic diffuse gamma-ray background (DGRB) refers to the sum-total of all gamma-ray radiation produced by DM annihilations throughout cosmic history \cite{ullio_cosmological_2002}. The amplitude of this signal depends on the large scale distribution of DM in the universe, the evolution of the isolated halo mass function, the concentration-mass relationship, and of course the density profile. Clumpy substructure may also play an important role, by boosting the annihilation luminosity of individual halos (see below). Large scale numerical simulations (e.g. Millennium-II \cite{boylan-kolchin_resolving_2009}) have been used to make predictions for both the amplitude of such a signal and the level of its anisotropies \cite{zavala_extragalactic_2010}, and these have been confronted with current Fermi data \cite{fornasa_characterization_2012}. The uncertainties of these constraints are dominated by the unknown contribution of subhalos below the simulations' resolution limit.

\item \textbf{Galactic Diffuse Gamma-ray Background}

   A second DGRB should arise from DM annihilations within the Milky Way's halo, with one component stemming from the smooth halo DM and another from clumpy substructure. The substructure component is expected to have a shallower angular intensity profile than the host halo component \cite{pieri_dark_2008,kuhlen_dark_2008}, for two reasons: (i) since it consists of the combined emission from a population of subhalos, it should scale with radius as the number density of subhalos $n_{\rm sub}(r)$, rather than as the square of the DM density $\rho_{\rm host}(r)^2$, and (ii) $n_{\rm sub}(r)$ is anti-biased with respect to $\rho_{\rm host}(r)$, with less substructure found close to the host's center. DM substructures introduce characteristic anisotropies in the Galactic DGRB \cite{siegal-gaskins_revealing_2008}, which may allow the signal to be disentangled from an astrophysical DGRB. 

   The detectability of the Galactic DGRB from DM annihilation thus depends on the abundance of substructure, its internal structure (concentrations and density profiles), and its spatial distribution within the host halo. If the substructure contribution remains sub-dominant, the shape of the Milky Way's DM halo may determine the large-scale angular variations of the signal.

\item \textbf{Galaxy Clusters}

  Galaxy clusters are the most massive gravitationally bound systems in the universe, and thus have long been considered good targets for indirect detection searches \cite{colafrancesco_multi-frequency_2006}. The detectability of an annihilation signal from clusters relies on a substantial cross section boost (either from substructure or Sommerfeld enhancement, see below) \cite{pinzke_prospects_2011,han_constraining_2012}, and the resulting emission would likely be extended. A difficulty is that any gamma-ray signal from annihilation has to compete with the cosmic ray induced gamma-ray emission \cite{jeltema_gamma_2009}. Nevertheless, H.E.S.S. \cite{abramowski_search_2012} and Fermi \cite{han_constraining_2012} data have been able to constrain DM parameters, and there is tentative evidence for a $\sim 130$ GeV line signal from a subset of the most promising cluster targets \cite{hektor_evidence_2012}.

\item \textbf{The Galactic Center}

  The most prominent DM annihilation signal is thought to arise from the Galactic Center (GC) \cite{gondolo_dark_1999}, given its proximity ($\sim8$ kpc) and the expected high DM density there. Unfortunately, the GC is also an extraordinarily astrophysically active place \cite{genzel_galactic_2010} hosting numerous SN remnants, pulsars, X-ray binaries, and other high-energy sources, not to mention a super-massive black hole. Although these astrophysical foregrounds encumber DM searches directed towards the GC, it nevertheless has remained a popular target for indirect detection efforts. In fact, several gamma-ray ``excesses'' and anomalies from the GC have been reported \cite{goodenough_possible_2009,hooper_dark_2011,hooper_origin_2011,abazajian_detection_2012}, including the recent intriguing reports of a gamma-ray line at $\sim 130$ GeV \cite{bringmann_fermi_2012,weniger_tentative_2012,tempel_fermi_2012,su_strong_2012}. It is too early to confidently claim a detection of DM annihilation for any of these signals, and additional data will be necessary before statistical fluctuations, instrumental effects, or astrophysical sources can be ruled out.

  The strength of the GC DM annihilation signal depends sensitively on the shape of the Milky Way host halo's DM density profile at scales that are currently not well resolved in numerical simulations. Predictions thus rely on extrapolations of fitting function that have been calibrated at larger radii (several hundred of pc) to a small number of high resolution simulations \cite{diemand_clumps_2008,stadel_quantifying_2009,navarro_diversity_2010}. Furthermore, the gravitational potential is baryon dominated at the GC, and one must thus account for modifications of the DM density profile due to baryonic physics. As discussed in more detail below (\S\ref{sec:DM+hydro}), these processes may lead to either a steepening or a flattening of the profile, and may even displace the location of maximum DM density from the dynamical center.

\item \textbf{Milky Way Dwarf Galaxy Satellites}

  The most DM dominated objects known are the Milky Way dwarf spheroidal satellite galaxies, which have mass-to-light ratios of up to a 1000 \cite{simon_kinematics_2007}. They are thus natural candidates for indirect detection searches \cite{strigari_most_2008}. Since their distances are fairly well known, the detectability of their DM annihilation signal is determined by the mass, concentration, and density profile of their DM host halos. For many of these systems, stellar kinematic data has provided tight constraints on the enclosed mass within the stellar half-light radius \cite{wolf_accurate_2010}, under assumptions of equilibrium and spherical symmetry. Fermi data from individual and stacked dwarf galaxies have provided some of the most stringent limits on the DM annihilation cross section, extending to below $3 \times 10^{-26} \, {\rm cm}^3 \, {\rm s}^{-1}$ for a DM particle mass of $\sim 40$ GeV and annihilation into pure $b\bar{b}$ \cite{ackermann_constraining_2011}, but these limits assume cuspy NFW-like DM density profiles and may be significantly weakened if baryonic processes or departures from the CDM assumption result in a flatter profile than inferred from DM-only simulations.

\item \textbf{Dark Subhalos}

  The vast majority of subhalos predicted in CDM cosmology are expected to be completely dark, since their masses are too low to have allowed gas to cool and form stars \cite{efstathiou_suppressing_1992}. Individual dark subhalos may be promising sources for indirect detection searches, and results from high-resolution simulations have been used to quantify their detectability \cite{baltz_pre-launch_2008,kuhlen_dark_2008,pieri_implications_2011}. The Fermi point source catalog contains hundreds of ``unassociated'' sources without identified astrophysical counterparts \cite{nolan_fermi_2012}, and it is possible that DM subhalos may be lurking among them \cite{belikov_searching_2012}. Very nearby sources could appear as faint, spatially extended gamma-ray sources to Fermi \cite{kuhlen_dark_2008}, and it may even one day be possible to measure proper motions of very nearby subhalos \cite{koushiappas_proper_2006}. 

Once again, these results are highly uncertain due to insufficient knowledge of the abundance, spatial distribution, and luminosity-mass relation of subhalos on scales below the simulations' resolution limit, as well as their ability to survive interactions with the Galactic disk.

\item \textbf{Local Anti-matter}

  DM annihilations in the local neighborhood would produce high energy positrons and anti-protons, either through direct annihilation into leptons ($e^-e^+, \mu^-\mu^+, \tau^-\tau^+$) or via the hadronization and decay of other annihilation products. These high energy anti-particles might be detectable as an excess over astrophysical cosmic ray backgrounds, and have been searched for by the Fermi \cite{abdo_measurement_2009}, H.E.S.S. \cite{aharonian_probing_2009}, PAMELA \cite{adriani_anomalous_2009}, ATIC-2 \cite{chang_excess_2008}, and AMS-02 \cite{pato_discriminating_2010} experiments, among others. Owing to energy losses from inverse Compton and synchrotron radiation, the propagation distance for positrons is short ($\sim 1$ kpc), and thus any injection of positrons from DM annihilations would have to originate from the local neighborhood. The expected contribution from DM annihilations hence depends on the local density of the Milky Way halo at 8 kpc. The presence of subhalos within $\sim 1$ kpc of Earth could affect both the amplitude of this signal and its energy spectrum \cite{kuhlen_atic_2009}. Improved numerical simulations with higher resolution and accounting for baryonic physics effects will be necessary to better characterize the role of local DM annihilations in the high energy cosmic ray spectrum.

\item \textbf{Neutrinos from Earth \& Sun}

  DM annihilations occurring in the center of the Sun or Earth could produce high energy neutrinos that may be observable with neutrino observatories like AMANDA \cite{achterberg_limits_2006} and IceCube \cite{abbasi_limits_2009}. DM particles can be captured by the Sun and Earth through elastic scattering off of heavy nuclei \cite{press_capture_1985}. Subsequent scatterings then thermalize the population of bound DM particles, and an equilibrium is established between annihilations and capture. The strength of the signal depends on the local DM density, but additionally also on the speed distribution of incident particles, since lower speed particles are easier to capture \cite{sivertsson_wimp_2012}. Rates are thus especially sensitive to the presence of a ``dark disk'' component \cite{bruch_dark_2009}, which can result in a marked increase in the fraction of DM particles traveling at low speeds with respect to the solar system.

\item \textbf{Substructure Boost Factors}

Given that the smallest collapsed structures in WIMP CDM are expected to be $\sim 10^{-12} - 10^{-3} \Msun$ \cite{profumo_what_2006,bringmann_particle_2009}, even the highest resolution numerical simulations can only resolve a small fraction of the expected substructure hierarchy. Since annihilation rates scale with the square of the density $\rho$ and $\langle \rho^2 \rangle \geq \langle \rho \rangle^2$, any unresolved small-scale clumpiness should result in a boost of the annihilation rate. Halo annihilation rates calculated from average density profiles (like NFW or Einasto), or even directly from the simulated particle distribution, could well underestimate the true luminosity by several orders of magnitude. This so-called substructure boost factor has been invoked to motivate effective annihilation cross sections orders of magnitude larger than the thermal relic value \cite[e.g.][]{baltz_cosmic_2002,hooper_can_2004,de_boer_evidence_2005,cumberbatch_local_2007,pinzke_prospects_2011}.

\begin{figure}
\centering
\includegraphics[width=\textwidth]{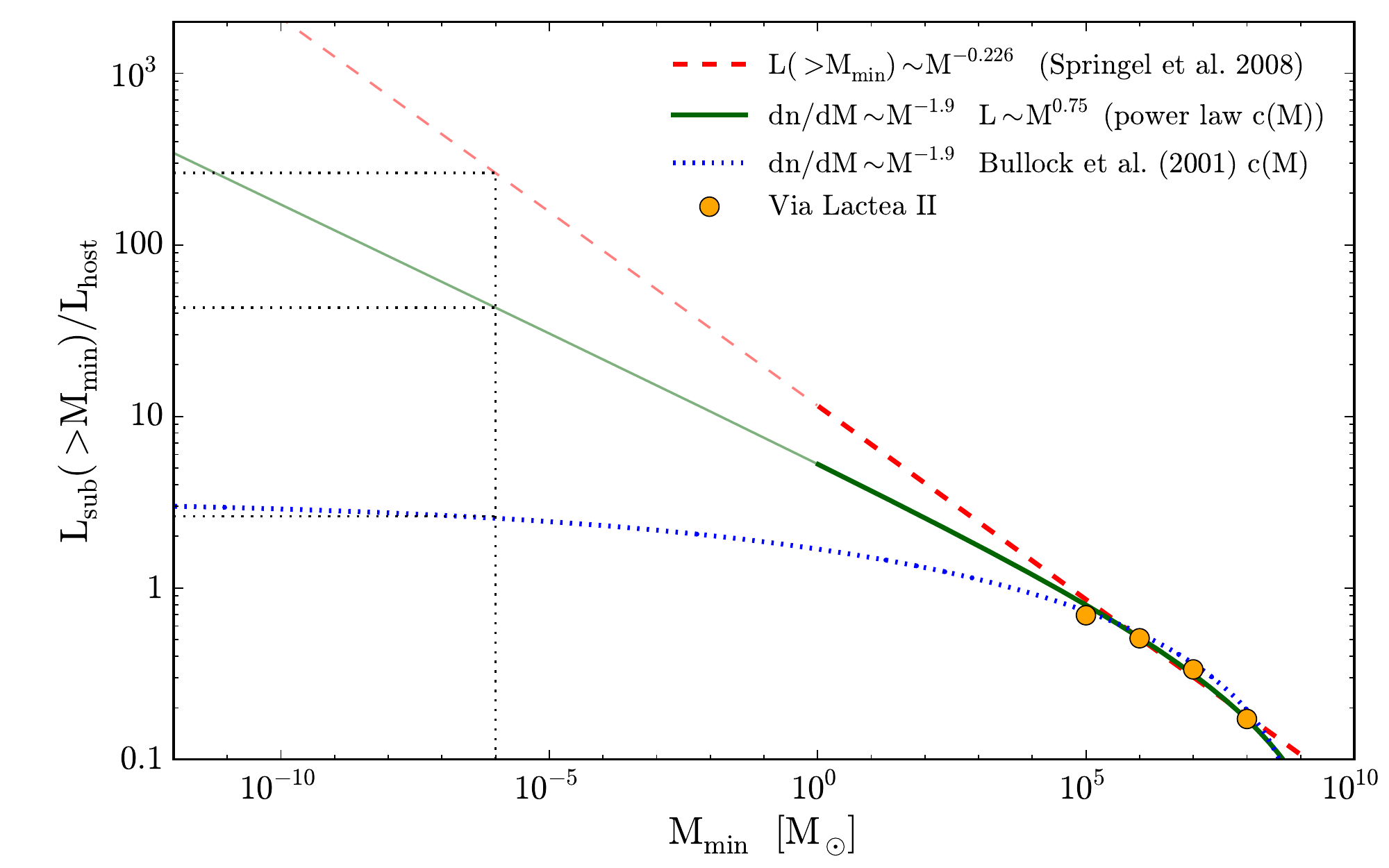}
\caption{An extrapolation of the subhalo contribution to the total
  luminosity to masses far below the simulation's resolution limit.
  Depending on what one assumes for the concentration-mass relation,
  one can get very different total substructure boost factors. Extrapolations from the high-mass behavior seen in simulations (red dashed) or assuming a constant power law concentration-mass relationship (green) are unlikely to hold at masses below $\sim 1 \Msun$ (visually indicated with thin faint lines).}
\label{fig:Lfraction}
\end{figure}

In the following we discuss two important facts about substructure boost factors that are perhaps not as widely appreciated as they should be:

\begin{enumerate}
\item There is no one single boost factor.

The expected substructure boost depends on the distance from the halo center, with results from state of the art simulations implying very little (or no) boost at the Galactic Center, possibly $\mathcal{O}(1)$ in the local neighborhood, and perhaps as large as 100 - 1000 for the total luminosity of a halo \cite{diemand_clumps_2008,springel_prospects_2008,kuhlen_dark_2008,kamionkowski_galactic_2010,pieri_implications_2011}. As a result a different boost factor applies to spatially extended sources (Galactic DGRB, MW satellite galaxies, dark subhalos) than for unresolved sources (distant halos, extra-galactic DGRB), and similarly a gamma-ray boost factor may not be the same as those for positron or anti-proton production \cite{lavalle_clumpiness_2007}. Furthermore, if a significant fraction of the mean density at a given radius is locked up in substructure, then properly accounting for the substructure boost will actually lower the smooth density contribution to the luminosity \cite{pieri_implications_2011}, further reducing the contrast between the outer regions of a halo and its center. The total halo luminosity boost likely depends on the mass of the halo, since numerical simulations indicate a roughly equal contribution from every decade of substructure mass, and larger mass host halos contain more decades of substructure mass \cite{pinzke_prospects_2011}. 

\item Substructure boosts depend \textit{sensitively} on subhalo properties many orders of magnitude below the resolution limit of state of the art simulations.

One approach to estimating the full substructure boost is to stay as close as possible to the results from ultra-high-resolution numerical simulations like Via Lactea II and Aquarius, by fitting the luminosity boost from all subhalos with mass greater than $M_{\rm min}$, $B(M_{\rm min}) = L(>M_{\rm min}) / L_{\rm smooth}$, to a power law of $M_{\rm min}$ over the 4-5 decades of substructure mass that are currently resolved in the simulations, and then extrapolating this power law down to the free-streaming cutoff scale. This approach was taken, for example, by Springel et al.~(2008) \cite{springel_prospects_2008}, who found $B(M_{\rm min}) \sim M_{\rm min}^{-0.226}$, and inferred a total boost factor for a Milky-Way-like halo of $230$ for $M_{\rm min} = 10^{-6} \Msun$.

Another approach is to use the numerical simulation results only to constrain the mass function of subhalos, which is measured to be a power law, $dn/dM_{\rm sub} \sim M_{\rm sub}^\alpha$ with logarithmic slope $\alpha \simeq -1.9$ \cite{diemand_clumps_2008,springel_aquarius_2008}, and to use an analytical approach to determine the subhalo luminosity-mass relation down to the smallest mass halos \cite{strigari_precise_2007,pieri_dark_2008,kuhlen_dark_2008}. The luminosity of a subhalo of mass $M$ is completely determined by its concentration $c$, $L/M \sim c^3 / f(c)$, where $f(c)$ depends on the shape of the density profile: for an NFW profile, $L/M$ scales approximately as $c^{2.24}$; for an Einasto profile, as $c^{2.46}$ \cite{kuhlen_dark_2010-1}. The subhalo annihilation luminosity-mass relation is then completely determined by the concentration-mass relation. Again, one may choose to use a simple power law relation, for example $c(M) \sim M^{-0.11}$, which well describes the concentration-mass relation of Galactic scale halos \cite{bullock_profiles_2001}. Alternatively one may choose a model in which the concentration of a halo reflects the mean density of the universe at its typical collapse time, as in the analytical model of Bullock et al. (2001) \cite{bullock_profiles_2001}. In this case, the concentration-mass relation is not a simple power law, but instead rolls over at low masses, and concentrations asymptotically become independent of mass. A comparison of the three approaches discussed so far is shown in Fig.~\ref{fig:Lfraction}, which demonstrates how sensitively the total halo boost factor depends on assumptions about the small scale behavior of subhalo luminosities. Depending on what one assumes for the concentration-mass relation, the total boost of a Milky Way halo ranges from 3 to 300 (for $M_{\rm min} = 10^{-6} \Msun$). Note that these three different approaches are not all equally likely to apply in reality. Simple extrapolations from the high-mass behavior observed in simulations or assuming a simple power law concentration-mass relation are inconsistent with expectation from theoretical models of CDM structure formation. Microhalo simulations find concentrations of the smallest and earliest collapsing DM halos that are incompatible with a single power law $c(M)$ over the full substructure hierarchy \cite{diemand_earth-mass_2005,diemand_early_2006,ishiyama_gamma-ray_2010}. Consequently, substructure boost factors much in excess of $\sim 10$ are unlikely to apply in nature.

A third approach, employed by Kamionkowski et al. (2010) \cite{kamionkowski_galactic_2010}, it to model the volumetric probability function of density fluctuations, calibrate it to a high resolution simulation (Via Lactea II) as a function of halo-centric radius, and then to integrate this PDF to obtain an estimate of the boost factor as a function of radius. This approach also results in a modest total boost factor for a galaxy-scale halo of $\mathcal{O}(10)$.
 
\end{enumerate}

Lastly, the annihilation luminosity can in principle also be enhanced by caustics in the fine-grained substructure, however recent numerical studies of this caustic boost find only percent level increases due to very efficient mixing in phase-space \cite{vogelsberger_streams_2011}.

\item \textbf{Sommerfeld Boost Factor}

  A second type of boost factor is of particle physics nature. When the mass of the force carrier particle is sufficiently lighter than the DM particle, the so-called Sommerfeld effect, a non-perturbative correction for long range attractive forces, can lead to a velocity-dependent enhancement in the annihilation cross section \cite{hisano_explosive_2004,arkani-hamed_theory_2009}. Instead of the usual $\langle \sigma v \rangle =$ constant behavior, with Sommerfeld enhancement the cross section scales as $\langle \sigma v \rangle \sim 1/v$, or even $1/v^2$ at resonances \cite{lattanzi_can_2009}. Although the effect typically saturates at small velocities ($v/c \sim 10^{-4} - 10^{-5}$) owing to the finite range of the interaction, this effect may significantly enhance the annihilation rate in subhalos compared to the smooth host halo, given the subhalos' lower velocity dispersions \cite{bovy_substructure_2009,robertson_dark_2009,kuhlen_exploring_2009}. The details depend on the predictions of numerical simulations of the velocity structure in the host and subhalos. On the fine-grained level of DM the Sommerfeld effect can have interesting implications. Whereas in non-Sommerfeld models the largest annihilation signal is expected to occur
near caustics due to their high density, this situation changes if Sommerfeld enhancement processes are invoked. In that case, cold low-velocity dispersion phase-space structures are enhanced compared to hotter
regions. Liouville's theorem dictates that DM is very hot in caustics to preserve the fine-grained phase-space density. Depending on the details of the Sommerfeld model this can make fine-grained streams more prominent for annihilation radiation than caustics, because streams are very cold due to their potentially low density. This can cause the  annihilation rate in streams to dominate
over the rate of the smooth mean density contribution in halos \cite{zavala_cosmic_2011}.

\end{enumerate}

\subsubsection{Direct Detection}

Direct detection refers to efforts to detect nuclear recoil signatures produced in rare DM-nucleus scattering events in shielded underground detectors. A large number of experiments are currently pursuing this goal, and are utilizing a variety of different technologies and target materials (see \cite{gaitskell_direct_2004} for a review). The expected event rate and recoil spectrum depends on the mass of the target nuclei and of the DM particle, on the nature of the interaction (spin-dependent vs. spin-independent), the nuclear form factor, the local DM density $\rho_0$, and the Earth frame velocity distribution $f(\vec{v})$ of incident DM particles. Until recently, most event rate and parameter exclusion calculations assumed a simplified model of the local DM distribution, taking the local DM density to be $0.3-0.4$ GeV cm$^3$ and a Maxwellian speed distribution with a 1-D velocity dispersion of $\sigma = 155$ km/s (such that the most likely speed $v_0$ is equal to the Galactic rotation speed at the solar circle, $v_0 = 220$ km/s), and an escape speed of 550 km/s.

In recent years these assumptions have been directly confronted with the predictions from high resolution simulations like Via Lactea II and Aquarius. The large number of self-bound subhalos found to be orbiting in the Milky Way's potential raises the question of whether one might expect large fluctuations in the DM density at the solar radius. If the Earth happened to be passing through a subhalo, for example, the local density of DM might significantly exceed the mean value at 8 kpc. Analytical calculations \cite{kamionkowski_galactic_2008} and direct ``measurements'' in simulations \cite{vogelsberger_phase-space_2009,kamionkowski_galactic_2010} indicate that the volumetric probability distribution of over-densities $\delta=\rho/\langle \rho \rangle$ consists of a narrow log-normal reflecting variations in the smooth halo density and a $\mathcal{P}_V(\delta) \sim \delta^{-2}$ power law tail extending over several orders of magnitude before steepening to $\delta^{-4}$ at an overdensity corresponding to the mean density of the universe at the collapse time of the smallest halos. Barring dramatic changes in the abundance and internal properties of subhalos below the simulations' resolution limit, the normalization of the power law tail at 8 kpc appears to be too low to lead to a non-negligible chance of the Earth lying in a substantial overdensity. It thus seems safe to use the mean value of the DM density at 8 kpc in direct detection calculation. However, what that value is remains uncertain at least at the factor of two level, with recent studies finding values ranging from 0.2 to 0.85 GeV cm$^{-3}$ \cite{catena_novel_2010,weber_determination_2010,salucci_dark_2010,garbari_new_2012}.

The speed distribution is another matter. DM-only simulations have definitively demonstrated that $f(v)$ shows clear departures from a pure Maxwellian \cite{diemand_velocity_2004,wojtak_radial_2005,hansen_universal_2006,vogelsberger_phase-space_2009,kuhlen_dark_2010}, with the typical shape instead exhibiting a deficit near the peak and an excess at lower and higher velocities. This is a consequence of the non-Gaussian nature of the three velocity components aligned with the density ellipsoid, with the major axis component being platykurtic (broader than Gaussian) and the other two leptokurtic (narrower) \cite{vogelsberger_phase-space_2009}. In addition to these coarse departures from a Maxwellian, additional small scale structures are often visible in the high speed tail, in the form of broad bumps that are stable in both space and time \cite{hansen_universal_2006,vogelsberger_phase-space_2009,kuhlen_direct_2012} and occasionally as narrow spikes at discrete speeds indicating the presence of a tidal stream or subhalo in the sample volume \cite{vogelsberger_phase-space_2009,kuhlen_dark_2010}. The presence of a strong ``dark disk'' \cite{read_dark_2009} would change the relative proportion of high speed and low speed particles, which could affect scattering rates and the recoil spectrum.

The scattering rate is proportional to $\int_{v_{\rm min}}^\infty f(v)/v dv$, where $v_{\rm min}$ is the minimum speed that can result in a recoil of energy $E_R$, which for elastic scattering is given by $v_{\rm min} = \sqrt{m_N E_R / 2\mu^2}$. $m_N$ is the mass of the target nucleus and $\mu = m_N m_{\rm DM} / (m_N + m_{\rm DM})$ is the reduced mass. The smaller $m_N$, the lower the speeds that are required to produce a given recoil energy. This implies that experiments with different target materials and different recoil energy sensitivities probe different parts of the speed distribution. Likewise, the mass of the DM particle can affect what range of speeds an experiment is sensitive to. For very massive particles ($m_{\rm DM} \gg m_N$), the experiment becomes insensitive to $m_{\rm DM}$, but for so-called \textbf{``vanilla'' WIMP DM} with $m_{\rm DM} \approx 100$ GeV, the current experiments' $E_R$-thresholds of $\sim 10$ keV correspond to $v_{\rm min} \approx 150$ km/s. The scattering rate will thus be dominated by fairly low speed particles near the peak of $f(v)$ and below. In this case, experiments will not be able to see effects from the interesting velocity substructures (the bumps and occasional spikes) that occur primarily at high speeds. A strong dark disk, on the other hand, may boost event rates.

On the other hand, with \textbf{inelastic DM} \cite{smith_inelastic_2001}, for which the relation between $E_R$ and $v_{\rm min}$ can become inverted, or \textbf{light DM} \cite{bottino_light_2004} ($m_{\rm DM} \lesssim 10$ GeV) experiments would be sensitive only to much higher speed particles, $v_{\rm min} \gtrsim 400 - 500$ km/s. In this case, the departures from Maxwellian, both global and local ones, would perhaps be more important. They could alter the shape and extent of current parameter exclusion curves \cite{kuhlen_dark_2010}, potentially reconcile some (but not all) of the currently discrepant results from different experiments \cite{frandsen_resolving_2012,kelso_toward_2012}, change the phase and amplitude of the annual modulation signal \cite{kuhlen_dark_2010} and shift it to higher recoil energies \cite{kuhlen_direct_2012}. \textbf{Directionally sensitive} experiments \cite{sciolla_directional_2009} should be especially sensitive to velocity substructure, since they typically have high recoil energy thresholds ($\sim 50$ keV), implying a large $v_{\rm min}$. The majority of high recoil energy particles may in fact be coming from a hotspot significantly offset from the direction anti-parallel to Earth's motion through the halo \cite{kuhlen_dark_2010}, and debris flow particles can result in ring-like features in the arrival direction \cite{kuhlen_direct_2012,bozorgnia_ring-like_2012}.

Besides generic WIMPs, axions provide another interesting DM candidate. {\bf
Axions} were introduced to explain the absence in Nature of strong-CP (Charge
conjugation and Parity) violations \cite{peccei_CP_1977}.  They are
expected to be extremely weakly interacting with ordinary particles, so that
they never were in thermal equilibrium in the early Universe.  This implies
that axions can be very light ($\mu$eV range) and nevertheless form a cold
(non-relativistic) component of cosmic matter.  The cosmological axion
population formed out of equilibrium as a zero momentum Bose condensate leading
to a very small velocity dispersion.  In the absence of clustering their
present day velocity dispersion would be negligible ($\delta v \sim 10^{-17}c$
compared to $\delta v \sim 10^{-10}c$ for generic WIMPs) making them a good CDM
candidate. Axions can be detected through their conversion to microwave photons
in the presence of a strong magnetic field \cite{sikivie_experimental_1983}.
Galactic axions have non-relativistic velocities ($\beta=v/c \sim 10^{-3}$) and
the axion-to-photon conversion process conserves energy, so that the frequency
of converted photons can be written as:
\begin{equation}
\nu_a = \nu_a^0 + \Delta \nu_a = 
241.8 \left(\frac{m_a}{1 \mu {\rm eV}/c^2}\right)\left(1+\frac{1}{2} \beta^2\right) \rm{MHz} 
\end{equation}
where $m_a$ is the axion mass that lies between $10^{-6}$~eV/$c^2$ and
$10^{-3}$~eV/$c^2$. $5\mu$eV axions would therefore convert into $\nu_a^0 \cong
1200$~MHz photons with an upward spread of $\Delta \sim  \cong 2$~kHz  due to
their kinetic energy. An advantage of axion detection compared to WIMP searches
is the fact that it is directly sensitive to the energy rather than to the
integral over the velocity distribution. Narrow velocity streams can therefore
be more easily detected and lead to spikes in the axion detection spectrum.
In that case the fine-grained structure can be relevant for detection
experiments. A low number of fine-grained streams could potentially leave a
distinct imprint in velocity-sensitive detector signals. For example, recent
simulations \cite{vogelsberger_streams_2011} predict that the most massive
fine-grained streams should be observable with axion detectors like ADMX.

\subsection{Relevance for Dark Energy Studies}\label{sec:relevance_de}

One of the simplest astrophysical observations, galaxy imaging and the measurement of their redshifts and angular positions on the sky, has emerged as a very powerful
method to explore the nature of DE. For instance, from this data baryon
acoustic oscillations, redshift-space distortions, abundance of galaxy
clusters and weak gravitational lensing can be measured, all of which can put constrains on the properties
of DE. Consequently, several galaxy surveys (e.g. DES, BOSS, BigBOSS, LSST,
JPAS, Euclid) are planned or underway to exploit this fact. DM numerical
simulations have been crucial in this process.

There are three main areas in which DM-only simulations are essential for
the cosmological exploration of DE.  Firstly, DM simulations have
been used to quantify and understand the effects of various DE models on
structure formation in the Universe \cite[e.g.][]{grossi_impact_2009, jennings_simulations_2010, fontanot_semi-analytic_2012}. 
This allows one to identify possible ways
of constraining or ruling out some DE models. Secondly, DM simulations are the
most reliable way to assess potential systematic errors on modeling and
cosmological parameter extraction from different experiments \cite[e.g][]{bianchi_statistical_2012}. Since most of the
DE probes involve complex and nonlinear processes, an accurate modeling of the
signal and related uncertainties in a given experiment is of paramount
importance in the discovery of new physics.  Thirdly, numerical simulations can
be used to construct mock galaxy and cluster catalogs, which help in the design
and correct interpretation of surveys aiming to constrain the properties of DE \cite[e.g.][]{overzier_millennium_2012}.
Since current and future surveys cover large solid angles and probe redshifts
out to $z \approx 2$, simulations are forced to very large box sizes and high
particle counts, in order to model volumes comparable to the surveys while
simultaneously resolving structure down to the scale of individual galaxies. 

In the following we briefly describe four of the most established probes
aiming to constrain the properties of DE and highlight the role of numerical
simulations in their development.

\begin{enumerate}
\item \textbf{Baryon Acoustic Oscillations}

Before recombination, the coupling between free electrons and photons via
Thomson scattering resulted in a distinct pattern of oscillations in the baryon
and temperature power spectra. These BAOs were also imprinted in the late-time
total mass field (albeit with smaller amplitude) due to gravitational
interactions between baryons and DM. Thus BAOs are expected to be present in
the galaxy power spectrum, and could be used as a cosmic standard ruler
\cite{eisenstein_baryonic_1998,seo_probing_2003,blake_probing_2003,linder_baryon_2003}.  Currently, the
feature has indeed been detected at high significance in different galaxy
surveys: 2dFGRS \cite{cole_2df_2005}, SDSS \cite{eisenstein_detection_2005}, WiggleZ \cite{blake_wigglez_2011}, 6dfGS \cite{beutler_6df_2011} and BOSS \cite{anderson_clustering_2012}, placing constrains on cosmological
parameters \cite[e.g.][]{blake_wigglez_2011, sanchez_clustering_2012}.  Future surveys are
expected to significantly tighten these constraints, in particular those on the
DE equation of state, which could rule out some DE candidates. 

Large-scale DM simulations played an important role in this development. They
showed that BAOs survive the diffusing effects of nonlinear evolution and of
galaxy peculiar velocities, and that they should be detectable in biased tracers
of the density field \cite{meiksin_baryonic_1999,seo_baryonic_2005,springel_simulations_2005, angulo_constraints_2005}. 
At the same time, numerical simulations unveiled significant distortions in
the shape of the acoustic oscillations due to these effects \cite{seo_baryonic_2005,angulo_constraints_2005,springel_simulations_2005,angulo_detectability_2008, sanchez_what_2008,smith_motion_2008}
which would lead to a systematic error on measurements of the acoustic scale.
However, recently methods to remove this bias have been proposed
\cite{eisenstein_improving_2007,padmanabhan_calibrating_2009,mehta_galaxy_2011}, and numerical simulations
have been used to test their validity and performance. In the future,
specially-designed numerical simulations will help us to understand and model
better the impact of structure and galaxy formation on the observed BAO signal
in the clustering of galaxies.

\item \textbf{Redshift space distortions}

The redshift of a galaxy not only contains information about its cosmological
distance, but also about its peculiar velocity. This difference between angular
positions and redshifts creates an anisotropy in the observed two-dimensional
galaxy correlation function that can be used to establish the relation between
density and velocity fields in the Universe \cite{kaiser_clustering_1987}. These
redshift-space distortions (RSD)  have been historically used to constrain the
value of the matter density of the Universe \cite{cole_constraints_1995,peacock_measurement_2001}, but
recently have also been employed to constrain the gravity law \cite{guzzo_test_2008,samushia_interpreting_2012, blake_wigglez_2011, beutler_6df_2012}. Current measurements are consistent with General
Relativity (GR), but future surveys are expected to significantly improve the 
constraints. Hypothetical departures from GR could be related to the DE model. 

However, extracting this information is not trivial. Numerical simulations have
shown that linear perturbation applies only on extremely large scales ($>50$
Mpc/h) \cite{hatton_modelling_1998,angulo_detectability_2008,jennings_modelling_2011}.  Quasi-linear corrections and the
so-called ``finger-of-God'' (FoG) cannot be neglected on smaller scales. In particular,
FoGs are produced by the motions of galaxies inside DM halos (whose velocity is
comparable to bulk motions produced by large-scale density perturbations),
which introduces a strong damping in the galaxy clustering along the line of
sight.

Given the increasing number of Fourier modes, it is desirable to use RSD down
to the smallest possible scales. Observations are usually modeled as a
combination of  linear theory expectations plus a damping to account for FoG
\cite[e.g.][]{guzzo_test_2008}. However, numerical simulations have highlighted the
pitfalls of this approach and the systematic error that it would introduce for
future surveys \cite{kwan_mapping_2012,okumura_systematic_2011,samushia_interpreting_2012,bianchi_statistical_2012}.  This finding has
fueled the development of more accurate new estimators which can robustly use
the clustering information at smaller scales \cite{tinker_redshift-space_2007,reid_towards_2011}.  This
is another example of the importance of DM numerical simulations in the optimal
exploitation of observational datasets.

\item \textbf{Abundance of galaxy clusters}

The position of galaxies in an optical survey can also be used to identify
galaxy clusters.  The number of these objects a function of their mass is of
great interest because it contains information about the underlying probability
distribution function of density perturbations in the Universe. The evolution
of the mass function on group and cluster scales has indeed been used to
measure cosmological parameters, helping to break degeneracies in the
constraints from other probes \cite{henry_x-ray_2009,mantz_observed_2010,vikhlinin_chandra_2009,rozo_cosmological_2010,benson_cosmological_2011,tinker_cosmological_2012}.  In addition, the highest mass end is
sensitive to primordial non-Gaussianities and early DE, and thus the detection
of massive, high-redshift clusters has been used as evidence of their existence
\cite[e.g.][]{mullis_discovery_2005,hoyle_implications_2011,foley_discovery_2011,brodwin_spt-cl_2010,baldi_high-z_2011,menanteau_atacama_2012,hoyle_critical_2012}.  However, uncertainties in the mass estimation of clusters and the
respective Eddington bias have seriously hampered these measurements
\cite{mortonson_simultaneous_2011,hotchkiss_quantifying_2011}.

The most relevant aspect of DM simulations for this cosmological probe is the
halo mass function and its dependence on cosmology \cite[e.g.][]{reed_toward_2012}. This prediction is usually
parametrized in terms of the linearly extrapolated variance of the underlying
DM field \cite{jenkins_mass_2001,warren_precision_2006,reed_halo_2007,tinker_toward_2008}.  However,
recently numerical simulations have shown  evidence for dependencies with other
parameters \cite{tinker_void_2009}. This finding could seriously limit the maximum
performance of current approaches to cosmological parameter extraction using
clusters. For this reason, in the future numerical simulations will probably
play a direct role in the modeling of the observed abundance of clusters.

Another important aspect in this probe is in the characterization of the
performance of cluster finder algorithms. DM-only simulations can quantifying
the impact of projection effects, misidentification of the cluster's center, and
false detections \cite{cohn_issues_2009,rozo_stacked_2011,angulo_scaling_2012},
as well as the relation between cluster mass and observed richness or weak
lensing signal.  Hydrodynamical simulations have an analogous role for
experiments using X-rays or thermal Sunyaev-Zeldovich detected clusters.

\item \textbf{Weak lensing}

The light from high-redshift galaxies gets distorted by intervening mass before
it reaches us.  Deep gravitational potentials cause large distortions which can
split the image of a galaxy into multiple lensed images, in a phenomenon known
as strong gravitational lensing. This has been discussed in 2.2.1 and can be
used to probe the mass of DM halos and even the law of gravity and the
nature of DM. Smaller distortions in the shape (and size) of background galaxies
caused by the cosmic web are known as weak gravitational lensing. These changes
in the properties  of galaxies can be related to integrals of the DM mass power
spectrum and thus they can be used to reconstruct the full three-dimensional DM
density field. This allows measurements of the growth of structure, which can be used to constrain DE and modified gravity \cite{schmidt_weak_2008}.

The shear of galaxies has been detected statistically in many surveys
\cite{wittman_detection_2000,bacon_detection_2000,refregier_cosmic_2002,hoekstra_constraints_2002,jarvis_weak-lensing_2003,semboloni_cosmic_2006,huff_seeing_2011} and has been used  to place constraints on cosmological parameters \cite{van_waerbeke_likelihood_2002,pen_three-dimensional_2003,fu_very_2008}.  For these, the main ingredient is the dependence of the nonlinear
DM correlation function on cosmology, which is normally taken from fitting
formulae calibrated using predictions of DM simulations \cite{seljak_analytic_2000,peacock_non-linear_1996,peacock_halo_2000,smith_stable_2003,hilbert_ray-tracing_2009}. However, weak
lensing measurements are affected by many sources of
systematic errors: most importantly the intrinsic alignments in the shape of
galaxies caused by tidal forces \cite{mandelbaum_detection_2006}, as well as the PSF ellipticity caused by atmospheric distortions \cite{heymans_shear_2006}, among others. Over the last few years extensive studies of these effects have been carried out, with DM simulations helping to create synthetic data as well as constraining the impact and magnitude of intrinsic alignments.

The next generation of surveys are expected to be able to reduce systematic
effects drastically, and thus will require high-precision predictions of
the nonlinear DM power spectrum down to small scales. Since perturbation
theory approaches can provide a reasonable description only in the mildly
nonlinear regime, the necessary predictions and modeling of data will 
have to rely on DM numerical simulations, either directly or via
emulators. 

\end{enumerate}

\section{Current State of the Art}\label{sec:state-of-the-art}

In this section we present a late 2012 snapshot of the state of the art of cosmological numerical simulations, with a focus on runs with particular relevance to the DM and DE problems. We first discuss DM-only simulations (\S\ref{sec:DM-only}), which are mature, mostly computational resource limited, and have been pushed to extremely high resolution, and then DM+hydro simulations (\S\ref{sec:DM+hydro}), which are algorithmically more challenging, less well developed, limited to lower resolution, and do not yet produce robust or even converged results.

\subsection{Dissipationless Dark-Matter-only Simulations} \label{sec:DM-only}

\subsubsection{Numerical Techniques and Codes} \label{sec:DM_techniques}

Pure DM simulations take the ansatz of completely neglecting any dissipational baryonic physics and treat all matter as collisionless DM. The density field is sampled with discrete ``N-body'' particles, whose gravitational evolution is governed by the Poisson-Vlasov equations in a coordinate system that is co-moving with the mean expansion of the universe. The effects of DE are generally confined to the expansion history, i.e. the translation between cosmic time and
expansion scale factor \cite{klypin_halo_2003,dolag_numerical_2004,kuhlen_dark_2005}. Many different techniques have been developed to solve this set of
equations, and we refer the reader to \cite{kravtsov_adaptive_1997,bertschinger_simulations_1998,stadel_cosmological_2001,springel_cosmological_2005,trenti_n-body_2008} for extensive discussion. For the present purpose, it suffices to briefly describe two of the major techniques in use today.

One is the so-called tree code \cite{barnes_hierarchical_1986}, in which the particle distribution is organized in a hierarchical tree structure. Contributions to the gravitational potential from distant particles are approximated by the lowest order terms in a multipole expansion of the mass distribution at a coarse level of the tree. If accuracy requirements demand it, the tree is ``opened'' to a higher level and a more detailed particle distribution is accounted for. This method reduces the computational complexity of the N-body problem from $\mathcal{O}(N^2)$ to $\mathcal{O}(N \log N)$, with a well controlled error. A further improvement to $\mathcal{O}(N)$ scaling is possible through the use of the Fast Multipole Method (FFM) \cite{greengard_fast_1987}, in which forces are calculated between two tree nodes rather than between individual particles and nodes. In order to avoid unphysical hard scatterings between nearby particles (which are just tracers of an underlying continuous density field), gravitational interactions are ``softened'' on small scales \cite{dyer_softening_1993}, typically either with a Plummer or a cubic spline kernel. The force resolution of this method is then given by the softening length $\epsilon_{\rm soft}$, which in DM-only simulations is usually kept constant in co-moving coordinates. \textsc{Pkdgrav2} \cite{stadel_cosmological_2001} is a prominent example of a pure tree code, and it uses FFM.

The second commonly used N-body technique is the adaptive particle-mesh (PM) method, in which the particles are deposited onto an regular mesh to produce a density field. The mesh structure is often adaptively refined in high density regions demanding increased force accuracy. The gravitational potential is obtained via Fourier transform on the periodic root grid (coarsest level), and a multi-grid relaxation technique is used to evaluated it on the refined grids. This method also achieves $\mathcal{O}(N \log N)$ scaling, but here $N$ refers to the number of mesh cells, which is typically taken to be $2^3$ times the number of particles. No explicit force softening is necessary, since particles interact with each other through a mean field not individually, and the force resolution is effectively given by the cell spacing of the most refined mesh. Examples of pure adaptive-PM codes are \textsc{Art} \cite{kravtsov_adaptive_1997} and \textsc{Ramses} \cite{teyssier_cosmological_2002}.

One of the most widely used cosmological simulation codes is the hybrid Tree-PM code \textsc{Gadget} \cite{springel_cosmological_2005}, which uses the PM method to evaluate long range forces and the tree method for short range interactions. The \textsc{Gotpm} code \cite{dubinski_gotpm:_2004} is another example of such a hybrid.

The choice of gravitational softening length in cosmological simulations with tree codes is a contentious issue that has not been truly resolved. The difficulty arises because there are conflicting demands on the softening \cite{merritt_optimal_1996,dehnen_towards_2001}: on the one hand it is desirable to choose as small of a softening as allowed by computational resources (the smaller the softening, the shorter the time steps, and the more expensive the simulation), since it represents a distortion of the true gravitational potential and leads to overmerging \cite{moore_destruction_1996}. On the other hand, smaller softening results in stronger unphysical two-body relaxation effects, which can cause spurious heating as well as artificial fragmentation \cite{melott_demonstrating_1997,splinter_fundamental_1998}. Some studies have advocated for softening lengths no smaller than the mean particle separation in the initial conditions \cite{melott_demonstrating_1997,splinter_fundamental_1998}, while others have argued that it is sufficient to choose a softening that suppresses unphysical discreteness effects in collapsed region \cite{power_inner_2003,diemand_two-body_2004}. Cosmological zoom-in simulations (see \S\ref{sec:sims}) generally employ softenings significantly below the mean initial condition separation of high-resolution particles (e.g. ranging from 0.006 to 0.02 for the zoom-ins in Table~\ref{tab:DM_simulations}). As an aside, the need to avoid two-body relaxation effects is responsible for the slow $\approx N^{1/3}$ rate of convergence of halo density profiles \cite{diemand_two-body_2004}. Note that due to the hierarchical nature of collapse in CDM, no matter how many particles are used in a simulation, the first structures to collapse are always the smallest halos that are resolved with only a small number of particles and hence susceptible to 2-body relaxation effects.

\subsubsection{Simulations}\label{sec:sims}

Cosmological DM-only simulations can be divided into two types: (i) full-box and (ii) \mbox{zoom-in} simulations. The former resolve the entire computational domain with a single particle mass and force resolution, and typically cover a representative volume of the universe, with box sizes ranging from $\sim 100$ Mpc to tens of Gpc. They are generally focused on resolving the large scale structure of the universe and are most useful for statistical studies of DM halos. We refer to this class as \textbf{cosmic scale simulations}. 

\begin{table}
\singlespacing
\centering
\small
 
\renewcommand{\arraystretch}{1.2}

\begin{tabular}{cc|cccccc}

\multicolumn{8}{c}{\large DM-only simulations} \\

\midrule[1.5pt]

\multicolumn{1}{l}{\large \textsc{Cosmic}} & \multicolumn{7}{c}{} \\

Name & Code & $\rm L_{box}$      & $\rm N_{p}$     & $\rm m_{p}$          & $\rm \epsilon_{soft}$ & $\rm N_{halo}^{>100p}$ & ref.\\
     &      & $\rm [h^{-1} Mpc]$ &$ [10^9]$ & $\rm [h^{-1} \Msun]$ & $\rm [h^{-1} kpc]$    & $[10^6]$          &     \\ \hline

  DEUS FUR & \textsc{Ramses-Deus} & 21000 & $550$ & $1.2 \times 10^{12}$ & 40.0$^\dagger$ & $145$ & \cite{alimi_deus_2012} \\  

  Horizon Run 3 & \textsc{Gotpm} & 10815 & 370 & $2.5 \times 10^{11}$ & 150.0 & $\sim\!190$ & \cite{kim_new_2011} \\

  Millennium-XXL & \textsc{Gadget-3} & 3000 & $300$ & $6.2 \times 10^9$ & 10.0 & 170 & \cite{angulo_scaling_2012} \\

        Horizon-4$\Pi$ & \textsc{Ramses} & 2000 & 69 & $7.8 \times 10^9$ & 7.6$^\dagger$ & $\sim\!40$ & \cite{teyssier_full-sky_2009} \\

   Millennium & \textsc{Gadget-2} & 500 & 10 & $ 8.6 \times 10^8$ & 5.0 & 4.5 & \cite{springel_simulations_2005} \\

   Millennium-II & \textsc{Gadget-3} & 100 & 10 & $6.9 \times 10^6$ & 1.0 & 2.3 & \cite{boylan-kolchin_resolving_2009} \\

 MultiDark Run1 & \textsc{Art} & 1000 &  8.6 & $8.7 \times 10^9$ & 7.6$^\dagger$ & 3.3 & \cite{prada_halo_2012} \\

        Bolshoi & \textsc{Art} & 250 & 8.6 & $1.4 \times 10^8$ & 1.0$^\dagger$ & 2.4 & \cite{klypin_dark_2011} \\


\multicolumn{8}{l}{$^\dagger$\footnotesize{For AMR simulations (\textsc{Ramses}, \textsc{Art}) $\epsilon_{\rm soft}$ refers to the highest resolution cell width.}} \\

\midrule[1.5pt]

\multicolumn{1}{l}{\large \textsc{Cluster}} & \multicolumn{7}{c}{} \\

Name & Code & $\rm L_{hires}$    & $\rm N_{p,hires}$ & $\rm m_{p,hires}$        & $\rm \epsilon_{soft}$ & $\rm N_{sub}^{>100p}$ & ref. \\
     &      & $\rm [h^{-1} Mpc]$ & $[10^9]$ & $\rm [h^{-1} \Msun]$ & $\rm [h^{-1} kpc]$    &  $[10^3]$     & \\ \hline

Phoenix A-1 & \textsc{Gadget-3} & 41.2 & 4.1 & $6.4 \times 10^5$ & 0.15 & 60 & \cite{gao_phoenix_2012} \\
\midrule[1.5pt]

\multicolumn{1}{l}{\large \textsc{Galactic}} & \multicolumn{7}{c}{} \\

Name & Code & $\rm L_{hires}$ & $\rm N_{p,hires}$ & $\rm m_{p,hires}$ & $\rm \epsilon_{soft}$ & $\rm N_{sub}^{>100p}$ & ref. \\
     &      & $\rm [Mpc]$    & $[10^9]$ & $\rm [\Msun]$  & $\rm [pc]$           &  $[10^3]$   & \\ \hline

 Aquarius A-1 & \textsc{Gadget-3} & 5.9 & $4.3 \times 10^9$ & $1.7 \times 10^3$ & 20.5 & 82 & \cite{springel_aquarius_2008} \\

        GHalo & \textsc{Pkdgrav2} & 3.89 & $2.1 \times 10^9$ & $1.0 \times 10^3$ & 61.0 & 43 & \cite{stadel_quantifying_2009} \\

Via Lactea II & \textsc{Pkdgrav2} & 4.86 & $1.0 \times 10^9$ & $4.1 \times 10^3$ & 40.0 & 13 & \cite{diemand_clumps_2008} \\

\midrule[1.5pt]

\end{tabular}

\caption{Current state of the art in DM-only simulations on cosmic, cluster, and galactic scale, ordered by number of simulation particles. $\rm L_{hires}$ is a proxy for the size of the high-resolution region in zoom-in simulations, and is defined to be equal to the size of a cube at mean density enclosing all high resolution particles. $\rm N_{halo/sub}^{>100p}$ is the number of halos in the box (\textsc{Cosmic}) or subhalos within $r_{50}$ (\textsc{Cluster} and \textsc{Galactic}) with at least 100 particles at $z=0$. In some cases (DEUS FUR, Horizon-4$\Pi$) mass functions have not been published, and so we estimated $\rm N_{halo}^{>100p}$ from a Sheth \& Tormen \cite{sheth_large-scale_1999} mass function fit.}
\label{tab:DM_simulations}
\end{table}

\begin{table}
\singlespacing
\centering
\scriptsize
 
\renewcommand{\arraystretch}{1.2}
\renewcommand{\tabcolsep}{0.025in}

\begin{tabular}{lC{0.7in}cC{1in}cccccc}
\midrule[1.5pt]

Simulation & Supercomputer & Type & Center & Country & core-hours & $N_{\rm cores}$ & memory & disk space \\
           &               &      &        &         & $[10^6]$  &                & [TB]   & [TB] \\ \hline

       DEUS FUR & \textit{Curie Thin Nodes} & Bullx B510 & Tr\`es Grand Centre de Calcul (TGCC) & France & 10 & 38016 & 230 & 3000 \\

  Horizon Run 3 &  \textit{Tachyon II} & Sun Blades B6275 & KISTI Supercomputing Center & Korea & 4 & 8240 & 21 & 400 \\

 Millennium-XXL &  \textit{JuRoPa} & Bull / Sun Blades & Forschungzentrum J\"ulich & Germany & 2.86 & 12288 & 28.5 & 100 \\

 Horizon-4$\Pi$ & \textit{Platine} & Bull Novascale 3045 & Commissariat a l'Energie Atomique & France & 8 & 6144 & 14.7 & 300 \\

Millennium & \textit{p690} & IBM Power 4 & Rechenzentrum Garching & Germany & 0.35 & 512 & 1 & 20 \\

  Millennium-II & \textit{VIP} & IBM Power 6 & Rechenzentrum Garching & Germany & 1.4 & 2048 & 8 & 35 \\

 MultiDark Run1 & \textit{Pleiades} & SGI Altix ICE & NASA Ames Research Center & USA & 0.4 & 4000 & 8 & 20 \\

        Bolshoi & \textit{Pleiades} & SGI Altix ICE & NASA Ames Research Center & USA & 6 & 13900 & 12 & 100 \\

    Phoenix A-1 & \textit{DeepComp 7000} & HS21/x3950 Cluster & Chinese Academy of Science & China & 1.9 & 1024 & 3 & 15 \\

   Aquarius A-1 & \textit{HLRB-II} & SGI Altix 4700 & Leibniz Rechenzentrum Garching & Germany & 3.5 & 1024 & 3 & 45 \\

          GHalo & \textit{Marenostrum} & IBM JS21 Blades & Barcelona Supercomputing Center & Spain & 2 & 1000 & 1 & 60 \\

  Via Lactea II & \textit{Jaguar} & Cray XT4 & Oak Ridge National Lab & USA & 1.5 & 3000 & 0.3 & 20 \\

\midrule[1.5pt]

\end{tabular}

\caption{Supercomputers and computational resources utilized for each simulation.}
\label{tab:resources}
\end{table}

In the zoom-in class, simulations forgo capturing a representative fraction of the universe, and instead focus all available computational resources on one halo of interest, resolving its internal structure and substructure at the highest possible resolution. In order to achieve this goal, these simulations make use of nested initial conditions, in which the great majority of the computational domain is sampled only with very coarse resolution (large particle masses and force softenings), but a small volume containing an object of interest is resolved with much higher resolution. We distinguish between \textbf{cluster scale simulations}, in which the object of interest is a single galaxy cluster ($10^{14} - >\!10^{15} \Msun$), and \textbf{galactic scale simulations}, which zoom in on a single galactic halo ($\lesssim {\rm few} \times 10^{12} \Msun$). In both cases the halo of interest is typically identified in the $z=0$ output of a lower resolution full-box simulation. The particles contained within some multiple (usually $3-5$) of its virial radius are traced back to the initial conditions, and the low resolution particles in this Lagrangian volume are replaced with a nested set of higher resolution (lower mass) particles. The phases and amplitudes of the large wavelength Fourier modes used to calculate the initial condition displacements of these higher resolution particles are kept the same as in the coarse realization, but additional random modes are introduced at smaller wavelengths. This process ensures that the large scale matter distribution remains identical to the coarse simulation, but with greatly enhanced power in small scale substructure fluctuations. The technical details of this procedure are discussed in detail in the literature \cite{tormen_structure_1997,bertschinger_multiscale_2001,gao_early_2005,hahn_multi-scale_2011}.

In the following we briefly go over each of these three classes of simulations (cosmic, cluster, and galactic), highlighting both state of the art achievements as well as limitations and shortcomings. Over the last decade progress in this field has been driven by advances in computing technology and available resources at national supercomputing facilities, and were aided by the algorithmic developments discussed in the previous section. The simulations discussed below all required multiple millions of CPU-hours on thousands of processors, and required terabytes of memory and petabytes of disk storage. Some of the characteristics of these simulations are summarized in Tables~\ref{tab:DM_simulations} and \ref{tab:resources}, and visualizations for a subset are shown in Fig.~\ref{fig:4panel}.

\begin{enumerate}[i)]

\item \textbf{Cosmic scale}

In this class the state of the art has reached $\gtrsim 10$ billion particle simulations, with the current record holder (in terms of particle number), the recently completed DEUS Full Universe Simulation (FUR) \cite{alimi_deus_2012}, utilizing more than half a trillion particles in a 21 $h^{-1}$ Gpc box, which corresponds to the entire observable universe. It was run with a modified version of the \textsc{Ramses} code and took about 10 million CPU-hours using 38,016 MPI tasks on the European supercomputer \textit{Curie}. With a particle mass of $1.2 \times 10^{12} \Msun$ and force resolution of $\rm 40 \, h^{-1} kpc$, DEUS FUR cannot resolve individual galaxies and thus is limited to studying the large scale distribution of matter in the universe. The main driver of this simulation has been to quantify the imprint that DE leaves on cosmic structures (e.g. BAO), and how the nature of DE may be inferred from observations of large scale structure. In addition to a standard $\rm \Lambda$CDM ($w=-1$) run, two 
more FUR simulations at the same resolution but with different DE models ($w=-0.87$ Ratra-Peebles quintessence and $w=-1.2$ phantom fluid) have recently been completed.

Currently the only calculation able to simultaneously resolve scales relevant for BAO detection as well as those DM halos and subhalos expected to host galaxies to be seen in future surveys is the Millennium-XXL simulation. This simulation uses slightly fewer particles (303 billion) than DEUS FUR to represent the mass field in the Universe, but it has almost 200 times better mass resolution due to its smaller computational domain. It was run during the summer of 2010 at the J\"ulich Supercomputer Centre in Germany using 12,288 CPUs using a memory-efficient version of the \textsc{Gadget-3} code. The main goal of this simulation is to explore the impact of galaxy formation physics on cosmological probes, in particular for BAO detection and redshift-space distortion tests.

On considerably smaller but still cosmic scales, two of the most prominent simulations are the Millennium-II and the Bolshoi simulations. Millennium-II, a \textsc{Gadget-3} simulation, has 10 billion particles in a $\rm 100 \, h^{-1} Mpc$ box, for a particle mass of $6.9 \times 10^6 \Msun$. It cost 1.4 million CPU-hours on an IBM Power-6 supercomputer at the  Max-Planck Computing Center in Garching, Germany. Bolshoi, an \textsc{Art} simulation, uses 8.6 billion particles in a $\rm 250 \, h^{-1} Mpc$ box, giving a particle mass of $1.4 \times 10^8 \Msun$, and required 6 million CPU-hours on the \textit{Pleiades} supercomputer at NASA Ames. Both simulations have a force resolution of $\rm 1 h^{-1} kpc$. Although Bolshoi has 20 times poorer mass resolution, it covers 16 times more volume than Millennium-II. One additional difference between the two is the choice of cosmological parameters, with Millennium-II employing values inspired by the first year WMAP results ($\Omega_m = 0.25$, $\Omega_\Lambda = 0.75$, $h=0.73$, $\sigma_8 = 0.9$, and $n_s=1$), which for $\sigma_8$ and $n_s$ are more than $3 \sigma$ discrepant with the more recent WMAP 5-year and 7-year results, while Bolshoi used values ($\Omega_m = 0.27$, $\Omega_\Lambda = 0.73$, $h=0.70$, $\sigma_8 = 0.82$, and $n_s=0.95$) that are consistent with the more recent measurements.\footnote{Results from the Millennium simulations have been rescaled to the latest set
of cosmological parameters \cite{angulo_one_2010, guo_galaxy_2012}.} For both cases, the mass and force resolution is sufficient to resolve some of the internal (sub-)structure of Milky Way-like halos, while at the same time capturing a large enough sample of such galaxies ($\sim\!5000$ in Millennium-II, $\sim\!90,000$ in Bolshoi) to enable statistical studies. These simulations have provided precise and robust results on DM halo statistics like the mass function, subhalo abundance, mass and environment dependence of collapse times, and spatial correlation functions, spanning a wide range of scales, from dwarf galaxy halos to rich galaxy clusters.

\begin{figure}
\centering
\includegraphics[width=\textwidth]{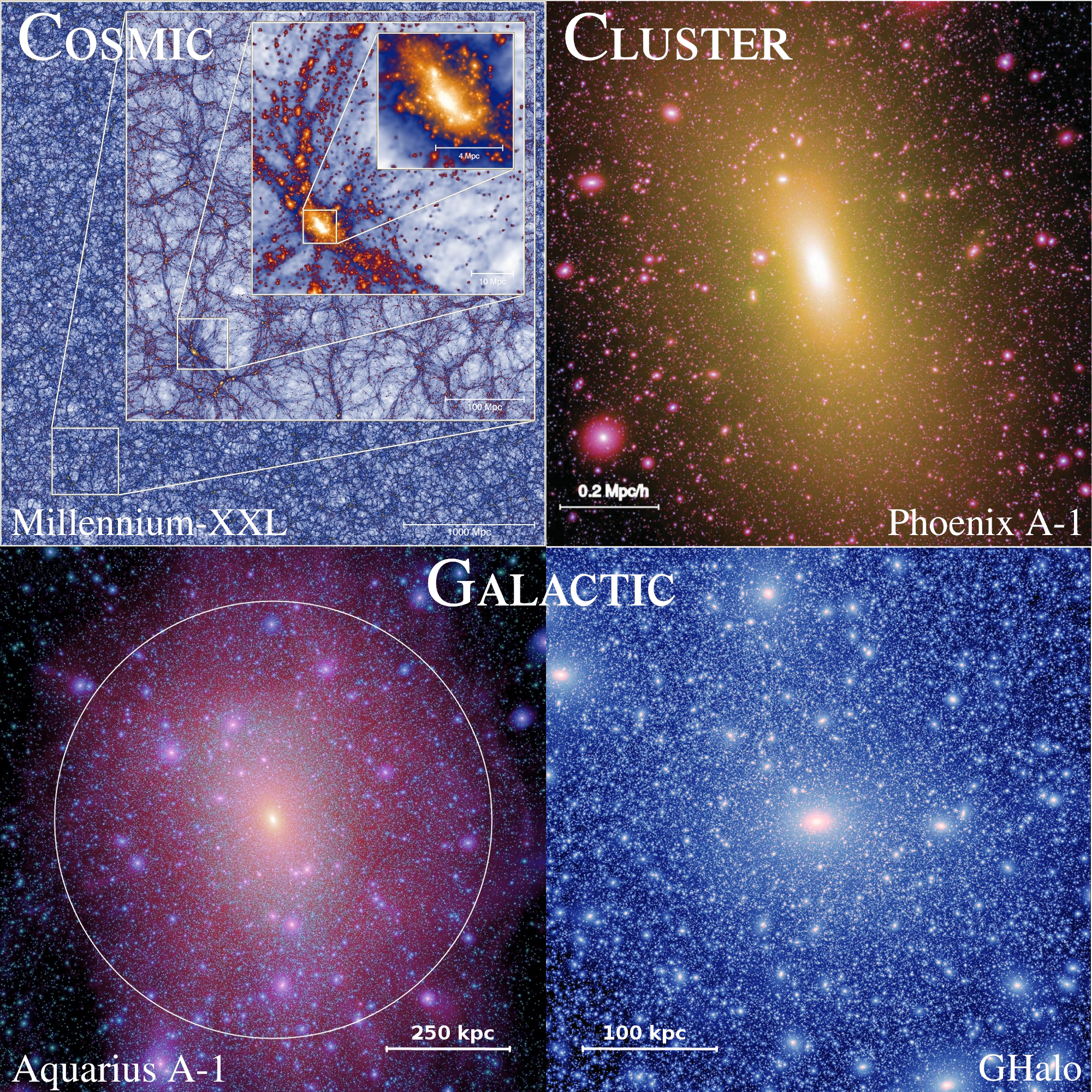}
\caption{Visualizations of state of the art simulations on cosmic (Millennium-XXL \cite{angulo_scaling_2012}, upper left), cluster (Phoenix A-1 \cite{gao_phoenix_2012}, upper right), and galactic scale (Aquarius A-1 \cite{springel_aquarius_2008}, lower left, and GHalo \cite{stadel_quantifying_2009}, lower right).}
\label{fig:4panel}
\end{figure}

\item \textbf{Cluster scale}

About ten years ago, cluster scale DM-only simulation were leading the effort to study the properties of individual DM halos and the abundance and properties of their substructure \cite{diemand_velocity_2004,reed_evolution_2005}. With around ten million high resolution particles, these simulations resolved thousands of subhalos and established important substructure scaling relations. DM substructure studies then shifted focus to the Galactic scale (see below), and until very recently cluster simulations had not pushed into the billion particle regime. The Phoenix simulation suite \cite{gao_phoenix_2012} has now changed that, with their highest resolution \textsc{Gadget-3} simulation employing 4.1 billion particles to resolve a $6.6 \times 10^{14} \, h^{-1} \Msun$ cluster, and identifying a total of almost 200,000 individual subhalos (60,000 with more than 100 particles). This simulation was run on 1024 cores of the \textit{DeepComp 7000} supercomputer of the Chinese Academy of Science and cost 1.9 million CPU-hours. Additional simulations of rare and dynamically young objects like galaxy clusters will help to clarify to what degree the internal structure of DM halos and substructure scaling laws are universal and self-similar.

\item \textbf{Galactic scale}

On Galactic scales the three flag-ship simulations are Via Lactea II \cite{diemand_clumps_2008}, Aquarius A-1 \cite{springel_aquarius_2008}, and GHalo \cite{stadel_quantifying_2009}, in chronological order. Via Lactea II (1.5 million core-hours on Oak Ridge National Lab's \textit{Jaguar}), was the first simulation to use over a billion high resolution particles to resolve a single halo, Aquarius A-1 (3.5 million core-hours at the Leibniz Rechenzentrum in Garching, Germany) the first to have over a billion particles within the virial volume of the halo, and GHalo (2 million core-hours on \textit{Marenostrum} at the Barcelona Supercomputing Center, Spain) is currently the simulation with the highest mass resolution. With particle masses ranging from 1000 to 4100 $\Msun$ and force resolutions from 20 to 60 pc, these simulations are able to resolve in unprecedented detail the formation and accretion history of Milky Way-sized DM halos ($M \approx 10^{12} \Msun$), their inner density profiles, and the properties and survival of stripped subhalo cores, as well as tidal debris orbiting within these systems. Density profiles have been reliably measured to $\sim\!\!100$ pc, and the substructure hierarchy is resolved over five decades in mass, down to $\sim\!\!10^5 \Msun$ subhalos. The Aquarius project simulated an additional five halos at somewhat lower resolution (particle mass $\approx 10^4 \Msun$), which has enabled a valuable initial assessment of halo-to-halo scatter.

Even though the simulations were run with different codes (\textsc{Gadget-3} for Aquarius, \textsc{Pkdgrav2} for Via Lactea II and GHalo) and used somewhat different cosmological parameters (most notably $(\sigma_8,n_s) = (0.9,1.0)$ and $(0.74,0.95)$, respectively), the numerical results agree remarkably well with each other when scaled by the mass of the simulated host halo. Some disagreements persist, however, in the interpretation of these results, for example in the assessment of the relative detectability of the Galactic DGRB indirect detection signal and that from individual subhalos \cite{kuhlen_dark_2008,springel_prospects_2008}, and in the self-similarity of the (sub-)substructure population \cite{diemand_clumps_2008,springel_aquarius_2008}.

\end{enumerate}

As will have become clear from the previous sections, these state of the art DM-only simulations on cosmic, cluster, and galactic scales require truly massive computational efforts (see Table~\ref{tab:resources}). Note that computational demands do not scale solely with N-body particle count, but also sensitively depend on the degree to which the simulations resolve small scale structure and non-linear clustering, mostly because more time steps are required. Single halo zoom-in simulations also require more communication and it is more difficult to balance their memory and CPU requirements than for full-box single resolution runs. For this reason a galactic-scale simulation like Via Lactea II required about the same number of CPU-hours (about one million) as the cosmic-scale Millennium-II run, even though the latter employed 10 times more particles. In addition to high computational demands at run time, simulations at this level present enormous challenges for data transfer, storage, and analysis (see \S\ref{sec:processing}.ii). As detailed nicely in the DEUS FUR simulation paper \cite{alimi_deus_2012}, analyzing the simulation often requires computing resources comparable to running it.

\subsection{Small scale challenges for Cold Dark Matter}\label{sec:CDM_challenges}

Predictions from CDM simulations of the large scale distribution of DM, post-processed to include mock galaxy populations, agree remarkably well with the observed clustering of galaxies measured in modern surveys like the Sloan Digital Sky Survey (SDSS) \cite{springel_large-scale_2006}. Yet at smaller scales the agreement between CDM predictions and observations is not as good: the number of dwarf satellite galaxies observed to be orbiting our Milky Way (and our nearest neighbor galaxy, M31) is less than one would naively infer from the predictions of DM-only simulations in a CDM cosmology \cite{moore_dark_1999,klypin_where_1999}. The severity of this so-called \textbf{Missing Satellites Problem} has been reduced in recent years through the discovery of more than ten new ultra-faint dwarf satellites in the SDSS \cite[and references therein]{simon_kinematics_2007}, raising the possibility that hundreds more remain yet to be discovered \cite{tollerud_hundreds_2008}. Nevertheless, reconciling the steep slope of the DM subhalo mass function with the shallow faint end of the satellite luminosity functions remains a theoretical and computational puzzle.

A second major challenge to CDM is the \textbf{Cusp/Core Controversy} concerning the central slope of DM density profiles in low mass galaxies. Two-dimensional stellar and gaseous kinematic measurements in low surface brightness field galaxies \cite{gentile_cored_2004,kuzio_de_naray_mass_2008,de_blok_core-cusp_2010,oh_central_2011}, as well as chemo-dynamical measurements in at least two Milky Way dwarf satellite galaxies \cite{walker_method_2011}, imply that the slopes of the DM density profiles are shallower than the NFW slope of $\sim-1$ predicted by CDM simulations without baryons.\footnote{Cored DM distributions have also been inferred for more massive spiral galaxies \cite[e.g.][]{donato_constant_2009}, but given that these systems are strongly baryon dominated, this observation is not commonly considered a major challenge for CDM.}

Lastly and possibly connecting the previous two concerns, it has recently been pointed out \cite{boylan-kolchin_too_2011,boylan-kolchin_milky_2012} that there may be a problem in the abundance of even the most massive Galactic subhalos. Dubbed \textbf{Too Big To Fail}, this problem refers to the inference from stellar kinematic data that the central densities of the ``classical dwarfs'' (bright satellites with luminosities greater than $10^5 L_\odot$) are too low to be consistent with inhabiting the most massive subhalos predicted in the Via Lactea II and Aquarius simulations. The consequence being that either there exists a population of massive subhalos orbiting within the Milky Way's virial volume that have remained completely dark and devoid of stars, despite the fact that less massive subhalos manifestly were able to form galaxies,  or that the Via Lactea II and Aquarius halos are somehow not representative of our Milky Way. For example, if the mass of the Milky Way were a factor of two less than in these simulations (but still within the range allowed from observational constraints), then the number of discrepant (too dense) halos may be small enough to not be a major worry \cite{wang_missing_2012,purcell_bailing_2012}. Alternatively, some process not captured in the DM-only simulations could reduce the central densities in the most massive subhalos.

\subsection{Simulations with Departures from Collisionless Cold Dark Matter}\label{sec:departures}

The two assumptions that underlie all DM-only simulations described so far are (i) that DM is ``cold'', meaning that the cutoff in the density fluctuation power spectrum occurs on scales far below what is resolved in the simulations, and (ii) that it is collisionless, meaning that the only dynamically relevant interactions are gravitational, i.e. that any self-scattering effects are negligible. Although these assumptions are theoretically well motivated, holding for example for most supersymmetric DM models as well as for axions, they are not a priori requirements. A number of studies in the literature have investigated whether departures from the assumptions of cold and collisionless DM can provide solutions to the small scale challenges to CDM discussed in the previous section.

In the \textbf{Warm Dark Matter (WDM)} scenario the DM particle exhibits some non-negligible thermal velocities at high redshifts, instead of being truly cold. In this case, free streaming in the early universe will erase small scale density fluctuations, preventing the formation of low mass DM halos. For a WDM particle of mass $m_\chi$ and temperature $T_\chi$, a cutoff in the power spectrum then occurs \cite{bode_halo_2001,viel_constraining_2005} at a scale of
\begin{equation}
k_{\rm FS} \approx 50 \, {\rm Mpc}^{-1} \, \left( \f{m_\chi}{2 \, {\rm keV}} \right) \, \left( \f{T_\chi / T_\nu}{0.2} \right)^{-1}.
\end{equation}
Here $T_\chi$ has been expressed in terms of the temperature of the cosmic neutrino background $T_\nu$, and $(T_\chi/T_\nu)$ in general is model dependent. For thermal relic WDM particles (e.g. the gravitino \cite{pagels_supersymmetry_1982}) it can be related to the relic DM density via $\Omega_\chi h^2 = (T_\chi/T_\nu)^3 \, (m_\chi / 94 \, {\rm eV})$, but non-thermally produced particles (e.g. the sterile neutrino \cite{dodelson_sterile_1994}) can have a wide range of temperatures. Constraints from the Lyman-$\alpha$ forest limit the mass of a WDM particle to be $\gtrsim 2-8$ keV \cite{viel_constraining_2005,abazajian_constraints_2006,boyarsky_lyman-_2009}, depending on the details of the particle physics.

The suppression of small scale power may help to explain the puzzling dearth of Milky Way satellite galaxies \cite{maccio_how_2010,polisensky_constraints_2011}. A secondary effect arising from the lack of small scale structure is that the collapse times of halos above the free-streaming cutoff are delayed. This results in lower concentrations and reduced central densities, which may help to address the Too Big To Fail problem \cite{lovell_haloes_2012}. Low concentration halos are also more prone to tidal disruption, which further reduces the abundance of low mass objects in WDM halos. Lastly, we mention that WDM halos are expected to have central density cores, since the WDM particles' non-zero temperature results in a finite phase-space density in the early universe \cite{tremaine_dynamical_1979}, which by Liouville's theorem cannot grow during the formation of a halo. For realistic models that are consistent with constraints from the Lyman-$\alpha$ forest, however, it can be shown \cite{villaescusa-navarro_cores_2011,maccio_cores_2012} that phase-space density limited cores only occur on very small scales, $r_{\rm core}/R_{\rm vir} \lesssim 10^{-3}$, far below where there is observational evidence for a flattening of the DM density profile. WDM models by themselves thus do not appear to be capable of solving the cusp/core controversy.

There are several technical difficulties associated with numerically simulating WDM models. One is that the presence of a cutoff in the power spectrum in the initial conditions gives rise to the formation of a large number of spurious halos of purely numerical origin \cite{wang_discreteness_2007}. Another difficulty is that for sufficiently light WDM particles, small box sizes, or early simulation starting times the thermal velocities can become comparable to the bulk flows induced by the density fluctuations in the simulation's initial conditions. One should then apply thermal streaming velocities to the N-body particles, ideally by splitting each particle into $2N$ sub-particles and applying equal and opposite velocities randomly drawn from the primordial velocity distribution to each $N$ pairs of sub-particles \cite{klypin_structure_1993}. In practice, however, thermal streaming velocities can usually safely be neglected, unless one is simulating WDM with $m_\chi < 1$ keV (observationally ruled out) or using a boxsize smaller than 1 Mpc.

The hypothesis of essentially collisionless DM has also been contested. This leads to the idea of {\bf Self-Interacting Dark Matter (SIDM)} \cite{carlson_self-interacting_1992, machacek_self-interacting_1993, de_laix_constraints_1995, spergel_observational_2000}.  Initially SIDM models with a constant scattering cross section were quickly abandoned since those that could solve the small-scale CDM problems seemed to violate several astrophysical constraints, such as the observed ellipticity of the mass distribution of galaxy clusters \cite{miralda-escude_test_2002}  and the survivability of satellite halos \cite{gnedin_limits_2001}. But recently it was pointed out that some of these earlier constraints were overstated, and small velocity-independent self-interaction cross sections can have sizable effects on halo profiles without violating astrophysical constraints \cite{rocha_cosmological_2012}.  Also simple ad hoc velocity-dependent cross sections of the form $1/v^\alpha$ were explored \cite{colin_structure_2002}, yielding encouraging results that however lacked a proper underlying particle physics model.  More recently it was realized that self-interactions through a Yukawa potential can resolve the challenges facing velocity-independent SIDM models \cite{loeb_cores_2011}. The velocity dependence of scattering through the massive mediator of this dark force (similar to a screened Coulomb scattering in a plasma) could make scattering important for dwarf galaxies with low velocity dispersion, but unimportant at the much higher velocities encountered in galaxy clusters. Such models have been explored numerically and it has been shown that they can help to resolve some of the small-scale CDM problems through the formation of a central density core \cite{vogelsberger_subhaloes_2012}. Note that there also exist hybrid models (e.g. Atomic DM \cite{cyr-racine_cosmology_2012}, see Fig.~\ref{fig:Delta2}), in which the DM exhibits both self-interactions and suppression in small scale power.

\subsection{Simulations Including Baryons Physics}\label{sec:DM+hydro}

As we have seen, the tension on small scales between dwarf galaxy observations and the predictions of DM-only simulations might be an indication that the true properties of the DM particle differ from the cold and collisionless assumptions of these simulations. Unfortunately, however, the effects of modified DM particle physics can be mimicked by the complicated baryonic physics governing the formation of stars and galaxies inside DM halos. For this reason the problem of DM is closely coupled to the problem of galaxy formation, which of course is a worthy topic of study in its own right. A survey of the current state of the art in numerical simulations of galaxy formation is considerably beyond the scope of this review, and so in the following we instead provide a limited overview of recent results with particular pertinence to the DM and DE problems.

\subsubsection{Numerical Techniques and Codes}

The basic equations that are solved in cosmological hydrodynamics simulations are the Euler equations (conservation of mass, momentum, and energy) governing the flow of an ideal gas, coupled gravitationally to the DM sector through a source term in the energy equation and the Poisson equation. Neglecting viscosity (ideal gas) is a good assumption on cosmological and galactic scales, but ignoring the effects of magnetic fields and radiation less so, and accounting for magneto- and radiation-hydrodynamic effects in cosmological galaxy formation simulations is an active area of research \cite[e.g.][]{dolag_sph_1999,li_cosmomhd:_2008,wang_magnetohydrodynamic_2009,doumler_investigating_2010,wise_birth_2012}. Results from such studies, however, have not yet been brought to bear on the DM and DE problems, and so we focus here on the simpler pure hydrodynamic case.

Unlike for the purely gravitational N-body problem, where even conceptually quite different solvers (e.g. tree and adaptive PM codes, see \S\ref{sec:DM_techniques}) robustly produce similar results, the choice of method with which to treat the hydrodynamics can lead to marked differences in the results \cite{frenk_santa_1999,oshea_comparing_2005,agertz_fundamental_2007,tasker_test_2008,springel_e_2010,vogelsberger_moving_2011,kere_moving-mesh_2012}. Numerous detailed discussions of the different approaches and their relative advantages and disadvantages exist in the literature \cite[e.g.][]{trac_primer_2003,monaghan_smoothed_2005,agertz_fundamental_2007,springel_e_2010}, and we only briefly summarize the essentials here.

In general one can distinguish between Eulerian and Lagrangian methods, which discretize either space (Eulerian) or mass (Lagrangian). In \textbf{Smoothed Particle Hydrodynamics (SPH)} \cite{monaghan_smoothed_2005}, the most commonly used Lagrangian approach\footnote{Really it is ``pseudo-Lagrangian'', since shearing flows with distinct internal properties are not followed in a truly Lagrangian way on scales below the smoothing kernel \cite{vogelsberger_moving_2011}.}, the fluid flow is followed with particles, whose equations of motion are derived from a discretized particle Lagrangian \cite{springel_cosmological_2002}, which ensures excellent conservation of mass, momentum, energy, entropy, and angular momentum. Thermodynamic quantities (density, pressure, etc.) are obtained by smoothing over neighboring particles with a particle-dependent smoothing length. Advantages of SPH are that it is automatically adaptive, delivering higher resolution in collapsing regions, geometrically flexible, inherently Galilean invariant (errors don't depend on bulk flows), computationally inexpensive, and that it couples easily to an N-body gravity treatment (as for the DM). It is also often simpler to implement new physics prescriptions. However, SPH is not without its drawbacks. Its Lagrangian nature results in less resolution in lower density environments, and its estimates of thermodynamic quantities are noisy on the scale of the smoothing kernel. Artificial viscosity must be added in order to inject the entropy generated at shocks and to suppress unphysical oscillations in the states immediately surrounding it. This broadens the discontinuity to several smoothing lengths and has a tendency to make the method more dissipative. It has low accuracy for contact discontinuities, and as a result suppresses some astrophysically relevant fluid instabilities and mixing. However, not all of these disadvantages are inherent to the SPH method, and several recently proposed modifications have resulted in very promising improvements \cite{price_modelling_2008,cullen_inviscid_2010,read_sphs:_2012,hopkins_general_2012}. SPH simulations commonly employ a gravitational force softening length that is fixed in physical coordinates below some redshift (see \cite{scannapieco_aquila_2012}), which results in poorer spatial resolution at early times compared to a constant co-moving softening scale and may suppress early star formation. The most widely used SPH codes in the galaxy formation field are \textsc{Gadget} \cite{springel_cosmological_2005} and \textsc{Gasoline} \cite{wadsley_gasoline:_2004}.

With \textbf{Adaptive Mesh Refinement (AMR)}, the most widely used Eulerian hydrodynamics approach, the fluid flow is instead discretized in space. Euler's equations are solved on a regular mesh, which is adaptively refined in regions requiring higher accuracy \cite{berger_local_1989}. In finite volume methods, conserved quantities (mass, momentum, energy) are stored on the cells of the mesh, and their values are updated in a conservative fashion by solving for the fluxes across cell interfaces (for a primer, see \cite{trac_primer_2003}). In the widely used Godunov schemes, fluxes are calculated by considering the independent variables to be piecewise constant across each cell, with discontinuities at the cell boundaries, and solving the resulting Riemann problem for the characteristic waves (shocks, rarefactions, and contact discontinuities) traveling into the neighboring cells \cite{leveque_nonlinear_1998}. In practice higher order accurate methods (piecewise linear or piecewise parabolic) are commonly employed \cite{van_leer_towards_1979,harten_high_1983,colella_piecewise_1984}. Advantages of AMR are its low noise and high accuracy for shocks, contact discontinuities, and shear waves, allowing it to capture fluid instabilities with high fidelity, and its full control over where to place high resolution. Disadvantages are its higher algorithmic complexity, lower numerical stability, the fact that errors are not Galilean invariant, a tendency to overmix fluid, and that runtime memory requirements grow with refinement. The most commonly used AMR codes in the galaxy formation community are \textsc{Hydro-Art} \cite{kravtsov_adaptive_1997,kravtsov_origin_2003,rudd_effects_2008}, \textsc{Enzo} \cite{bryan_piecewise_1995,oshea_introducing_2004}, and \textsc{Ramses} \cite{teyssier_cosmological_2002} (with the \textsc{Flash} code \cite{fryxell_flash:_2000} about to join the fray \cite{mitchell_towards_2012}). 

\begin{figure}
\centering
\includegraphics[width=0.8\textwidth]{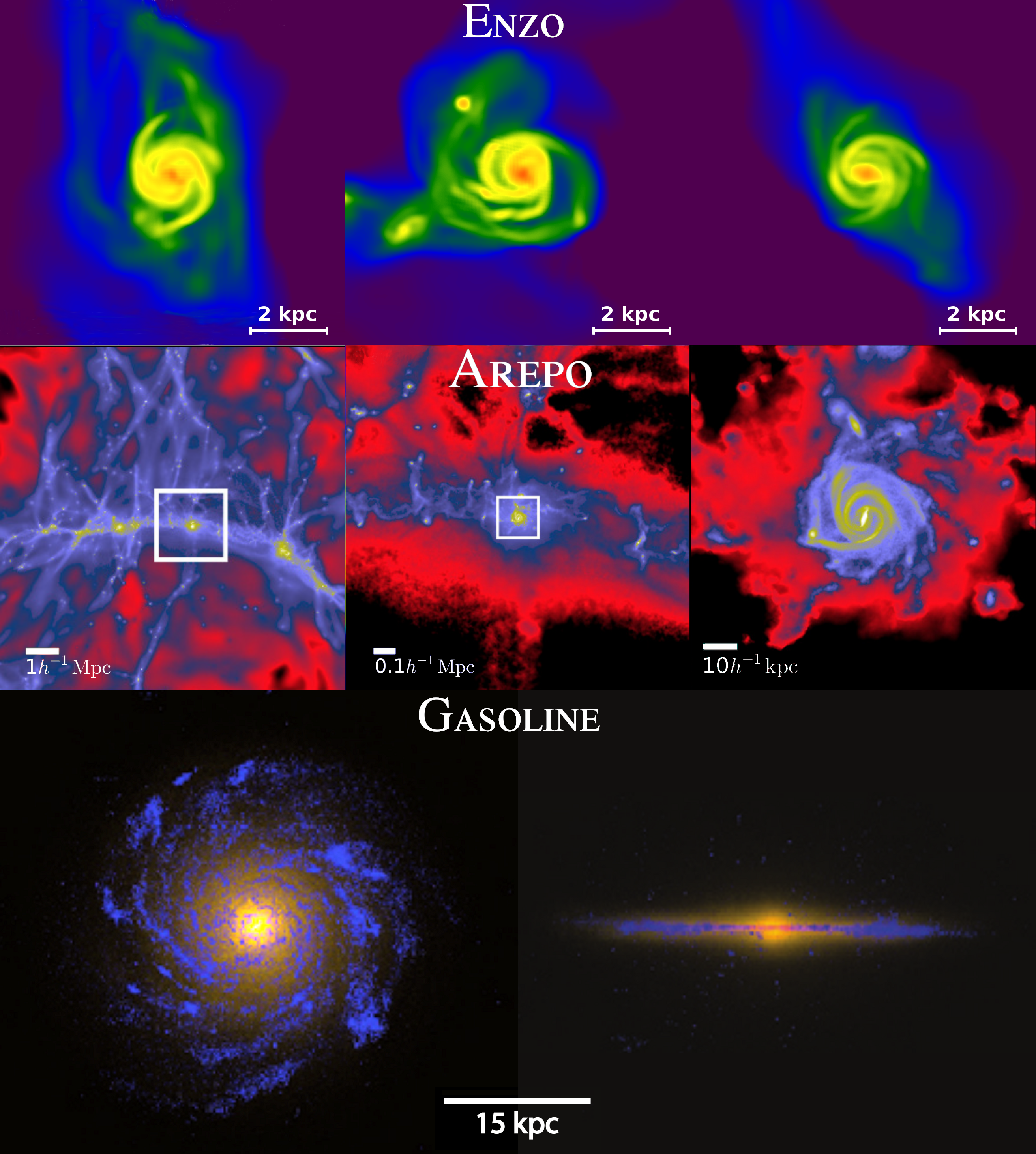}
\caption{Visualizations of three recent cosmological hydrodynamical galaxy formation simulations. Top row: gas surface density at $z=4$ in three galaxies (out of $\sim\!\!100$ in the box) simulated with the AMR code \textsc{Enzo} \cite{kuhlen_dwarf_2012}. Middle row: a series of zooms onto the density field surrounding a $z=2$ galaxy, simulated with the moving mesh code \textsc{Arepo} \cite{vogelsberger_moving_2011}. Bottom: Optical and UV composite images of Eris, a Milky Way-like galaxy simulated with the SPH code \textsc{Gasoline} \cite{guedes_forming_2011}.}
\label{fig:hydro}
\end{figure}

Both SPH and AMR techniques have weaknesses, which are related to the numerical
approach taken to solve the fluid equations. One advantage of SPH is its
pseudo-Lagrangian nature which fits very well the needs of cosmological
structure formation simulations, where adaptivity and a large spatial and
dynamical range is required. On the other hand AMR, as a finite volume scheme,
provides highly accurate results for fluid problems providing, for example,
very good resolution of shocks, discontinuities and mixing, which are typically
harder to resolve very well with SPH schemes. A natural way to combine the
advantages of SPH and AMR techniques is to allow for \textbf{Moving Meshes} in
the volume discretization. This idea goes back to the 1990's, where moving
meshes were first explored in the context of astrophysical applications
\cite{gnedin_softened_1995, pen_high-resolution_1998}. Although the idea of
having the computational mesh move with the hydrodynamical flow seems very
natural, its practical implementation turned out to be rather difficult.
Approaches relying on deformed Cartesian grids lead to problems in handling
grid deformation properly in fully astrophysical applications
\cite{gnedin_softened_1995,pen_high-resolution_1998}. Only recently new moving
mesh schemes have been developed, which are able to circumvent this problem
\cite{springel_e_2010, duffell_tess:_2011}.  These new schemes do not use
coordinate transformations like previous moving mesh codes in cosmology, but
instead employ an unstructured Voronoi tessellation of the computational domain.
The mesh-generating points of this tessellation are allowed to move freely,
offering significant flexibility for representing the geometry of the flow.  If
the mesh motion is tied to the gas flow, the results are Galilean-invariant
(like in SPH), while at the same time a high accuracy for shocks and contact
discontinuities is achieved (like in Eulerian schemes). Furthermore the mesh is
free of the distortion problems inherent to previous schemes. Using the \textsc{Arepo} code \cite{springel_e_2010}, this new
computational approach has recently been applied successfully to initial large-scale
cosmological simulations of galaxy formation \cite{vogelsberger_moving_2011,kere_moving-mesh_2012} (see Fig.~\ref{fig:hydro}).

An accurate treatment of the hydrodynamics is necessary, but far from sufficient. In order to model the galaxy formation process, simulations must go beyond adiabatic hydrodynamics and include \textbf{gas cooling} through radiative energy losses, as well as heating from an externally calculated meta-galactic UV background \cite{faucher-giguere_new_2009,haardt_radiative_2011}. The cooling is typically implemented through tabulated cooling functions, which provide externally calculated (using the Cloudy code \cite{ferland_cloudy_1998,smith_metal_2008,shen_circumgalactic_2012}, for example) equilibrium cooling rates as a function of density, ionization fraction, temperature, metallicity, and intensity of the UV background. Some codes follow the non-equilibrium abundance of hydrogen and helium species (including molecular hydrogen in some cases) coupled to gas cooling and heating by solving a chemical network sub-cycled on each hydrodynamic time step. The details of how gas cooling is implemented can make a difference in the simulation's outcome.

In regions that have cooled and condensed to sufficiently high density, stars will form. This process is captured with a sub-grid model, in which a fraction of the available gas is converted to ``star particles'' representing an entire stellar population, which from then on are treated as collisionless and evolved in the same fashion as DM particles. The nature of this \textbf{star formation} (SF) prescription is another important differentiating aspect of galaxy formation simulations. Most commonly, a Schmidt law \cite{schmidt_rate_1959} is employed, whereby the star formation rate (SFR) is proportional to the gas density divided by the local free fall time, resulting in a SFR proportional to $\rho^{3/2}$. However, some authors instead prefer a linear SF law corresponding to a fixed SF time scale \cite{kravtsov_origin_2003}, especially with simulations that distinguish between atomic and molecular gas phases and tie the star formation to the latter \cite{gnedin_modeling_2009,kuhlen_dwarf_2012,christensen_implementing_2012}. The SF efficiency is a free parameter in principle, but is usually set to a few percent, motivated by observational constraints \cite{krumholz_slow_2007}. Furthermore, a SF density threshold is often employed to limit SF to especially high density regions. Some methods additionally restrict SF to regions with a converging flow and a cooling time shorter than the dynamical time \cite{cen_cold_1993}. In order to prevent the formation of excessive numbers of star particles, a minimum star particle mass is sometimes enforced. For any given model, the parameters are typically tuned to reproduce macroscopic SF scaling laws like the Kennicutt-Schmidt relation \cite{kennicutt_global_1998}. Nevertheless, with so many different SF prescriptions and tunable parameters, it is not surprising that there is no unique solution, and that results vary greatly, especially in conditions far from where the prescriptions were calibrated.

Last, but by no means least, galaxy formation simulations must account for so-called \textbf{feedback} processes, which attempt to capture the injection of mass, momentum, energy, radiation, and metals (nucleosynthetic products) from massive young stars, evolved asymptotic giant branch stars, exploding supernovae (both type Ia and II), and accreting black holes into the surrounding interstellar medium. Feedback appears to be a crucial ingredient in explaining the macroscopic properties of individual galaxies \cite{ceverino_role_2009,sawala_formation_2010,governato_bulgeless_2010,guedes_forming_2011,stinson_making_2012} as well as galaxy population statistics \cite{schaye_physics_2010,dave_galaxy_2011,puchwein_shaping_2012}, and may be the key for explaining the small scale challenges CDM faces (see next section). At the limited resolution of current simulations, feedback is implemented using heuristic and often ad-hoc prescriptions (for a review, see \cite{thacker_implementing_2000}). Numerical difficulties that must be overcome include preventing the injected energy from immediately being lost owing to the high densities and cooling rates in star forming regions, accounting for radiation pressure from young massive stars, forming and maintaining large scale galactic outflows, and properly accounting for the mixing of metals. Even more so than for the SF prescription, results depend sensitively on the details of how feedback is implemented \cite{scannapieco_aquila_2012}, and this is probably the greatest source of uncertainty in present day galaxy formation simulations.

For completeness, we also briefly mention here that baryonic physics effects are also commonly accounted for through the use of Semi-Analytic Models (SAM). In this approach gas cooling, star formation, feedback from SNe and AGNs are all implemented in a simplified but self-consistent manner on top of halo matter merger trees derived from analytical calculations or from DM-only simulations. SAMs have been used to explore the connection between dark matter structure and galaxies (for a review, see \cite{baugh_primer_2006}) have helped to quantify how uncertainties in galaxy formation can propagate to DE studies \cite{angulo_detectability_2008,fontanot_semi-analytic_2012}.

\subsubsection{Baryonic effects on DM}

Given the uncertainties arising from the choice of hydrodynamic method, the sub-grid physics prescriptions, and the large number of adjustable parameters, it is not surprising that there is not yet consensus on how baryonic physics alters the distribution of DM in halos. In the following we report on some of the results from recent hydrodynamic galaxy formation simulations. We caution the reader, however, that in most cases these results should not be viewed as definitive answers, and that the conclusions are subject to change with higher resolution and improvements in the treatment of sub-grid physics prescriptions.

Concerning the large scale distribution of DM, it has been shown \cite{jing_influence_2006,rudd_effects_2008,van_daalen_effects_2011} that the inclusion of baryonic physics substantially alters the matter power spectrum on scales ($k \gtrsim 1 h \, {\rm Mpc}^{-1}$ or $l \gtrsim 800$) that are relevant for contemporary and future weak lensing galaxy surveys aiming to constrain the nature of DE. The weak lensing shear signal has also been shown to be affected by baryonic physics, especially when strong AGN feedback in galaxy clusters is accounted for \cite{semboloni_quantifying_2011}, and this can lead to significant biases (tens of percent) in the inferred DE equation of state parameter $w_0$. DM halo mass functions can also be affected, with baryons causing a $10\%$ enhancement in the cumulative mass function for $>10^{12} h^{-1} \Msun$ halos in one study \cite{rudd_effects_2008}, and $\sim 30$ percent deviations in the number density of $10^{14} h^{-1} \Msun$ halos in another \cite{stanek_effect_2009}, with the sign of the deviation (increase or decrease in halo mass) depending on how the baryonic physics was implemented. Note that $10-30 \%$ differences are larger than the 5\% statistical uncertainty in DM-only halo mass function \cite{tinker_toward_2008}. Gas mass fractions in galaxy clusters and halo mass scaling relations of the thermal Sunyaev-Zel'dovich effect depend on the treatment of baryons \cite{sembolini_music_2012}. Together these results imply that deep galaxy cluster surveys designed to tightly constrain cosmological parameters and the nature of DE must account for baryonic effects. 

The abundance and spatial distribution of subhalos inside galactic and cluster scale halos is another area likely affected by baryonic physics, and here again even the sign of the effect is unknown. Baryonic condensations within subhalos could increase the central density and make them more resiliant to tidal disruption. This could lead to an increase in the subhalo abundance close to the halo center, as seen in some simulations \cite{romano-diaz_dissecting_2009,romano-diaz_dissecting_2010}. On the other hand, if the host halo itself has a sizeable stellar disk, then disk shocking \cite{donghia_substructure_2010}, as well as interactions with individual stars \cite{goerdt_survival_2007}, could lead to enhanced subhalo destruction and lower their abundance in the inner part of the galaxy.

From the point of view of DM detection experiments, the most significant concern is the possibility that baryonic effects could modify the shape of the DM density profile. This is the arena of a long standing debate between advocates of adiabatic contraction increasing the central DM density on one side, and proponents of processes removing DM from the halo center on the other. If the cooling and condensation of gas proceeds slowly, with baryons gradually sinking to the center, then DM will be dragged in adiabatically, leading to a steepening of the DM density profile from the NFW slope of $\sim -1$ to something closer to isothermal $-2$. This \textbf{adiabatic contraction} effect was worked out analytically in \cite{blumenthal_contraction_1986} and its description has since been refined \cite{gnedin_response_2004,gnedin_halo_2011}. It is routinely observed in cosmological galaxy formation simulations \cite{romano-diaz_erasing_2008,johansson_gravitational_2009,abadi_galaxy-induced_2010,duffy_impact_2010,guedes_forming_2011,gnedin_halo_2011,zemp_impact_2012}.

If, on the other hand, baryonic material is rapidly delivered to the center, for example through cold flows \cite{kere_how_2005}, then adiabatic contraction may not operate and other dissipationless processes that transfer energy from the baryons to the DM could lower the central DM density \cite{lackner_dissipational_2010}. Examples of such \textbf{cusp-to-core conversion} processes are resonant interactions of the DM halo with a stellar bar\footnote{Some authors, however, instead find that the formation of a stellar bar actually increases the central DM density \cite{colin_bars_2006,dubinski_anatomy_2009}.} \cite{weinberg_bar-driven_2002,holley-bockelmann_bar-induced_2005,weinberg_bar-halo_2007}, the decay of a supermassive black hole binary \cite{milosavljevi_formation_2001}, dynamical friction of dense stellar clumps against the smooth background DM halo \cite{el-zant_dark_2001,tonini_angular_2006,romano-diaz_erasing_2008,romano-diaz_dissecting_2009}, and repeated and violent oscillations in the central potential due to energy injection from active galactic nuclei \cite{peirani_active_2008} or supernova-driven galactic outflows \cite{mashchenko_removal_2006,governato_cuspy_2012,pontzen_how_2012,teyssier_cusp-core_2012}. This last process in particular has recently attracted much attention, since galaxy formation simulations with very efficient ``blastwave'' supernova feedback have for the first time resulted in spiral galaxies with realistically low bulge-to-disk ratios \cite{governato_bulgeless_2010,guedes_forming_2011}, and also exhibit pronounced DM density cores \cite{brooks_why_2012}. The removal of the central DM density cusp through baryonic processes is especially interesting in the context of the Missing Satellites and Too Big To Fail problems of CDM (see \S\ref{sec:CDM_challenges}) \cite{zolotov_baryons_2012,brooks_baryonic_2012}.

Lastly, we mention the possibility that baryonic physics effects could lead to a displacement between the point of maximum DM density in a halo and its dynamical center \cite{mcmillan_halo_2005,kuhlen_off-center_2012}, as recently identified in the Eris simulation \cite{guedes_forming_2011}, one of the highest resolution and most realistic cosmological hydrodynamic simulations of a Milky Way-like galaxy (see Fig.~\ref{fig:hydro}). In this simulation, but not in comparable DM-only simulations, the DM offset was $\sim 340$ pc averaged over the last 8 Gyr, typically in the plane of the stellar disk, and aligned to about 30 degrees with the orientation of the stellar bar. Such an offset would considerably alter expectations for the indirect DM detection signal from the Galactic Center, where Sgr A*, the compact radio source associated with a supermassive black hole, is a good marker of the dynamical center of the Galaxy \cite{genzel_galactic_2010}. Intriguingly, the recently reported 130 GeV gamma-ray line from the Galactic Center \cite{bringmann_fermi_2012,weniger_tentative_2012,tempel_fermi_2012,su_strong_2012} appears to be offset by about 1.5 degrees (about 200 pc projected) from Sgr A*.

\section{The Next Decade}\label{sec:next_decade}

Having rextensively reviewed the current state of the art of cosmological DM and galaxy formation simulations, we now present our vision for this field for the next ten years. What progress should be possible and where should priorities be focused in order to maximize our understanding of the DM and DE problems?

\subsection{Dissipationless Dark-Matter-only Simulations}

On the cosmic scale, there is a need for very large numbers of high resolution simulations in large box sizes, scanning over cosmological parameters. Future DE surveys (e.g. DES, BigBOSS, LSST) will cover enormous volumes (10000's of square degrees out to $z \approx 2$) and are sensitive to quite low mass galaxies. In order to perform grid-based or Markov Chain Monte Carlo estimations of cosmological parameters, their errors, and especially the co-variances of their errors, it will be necessary to have highly accurate theoretical predictions of the non-linear clustering of matter for a finely sampled scan of cosmological parameters. At the desired percent level accuracy, such predictions can and must come from cosmological simulations \cite{heitmann_coyote_2010} -- hundreds to thousands of them \cite{sato_simulations_2009,takahashi_simulations_2009}. Each simulation must cover volumes comparable to the surveys, and have a mass and force resolution sufficient to resolve nonlinear clustering on galactic scales. Furthermore, owing to non-linear mode coupling, small scales are affected by the sampling variance of the largest modes, necessitating multiple realizations for any given cosmology. In the mildy non-linear regime ($k < 0.3 \, h \, {\rm Mpc}^{-1}$), it may be possible to greatly speed up cosmological parameter scans by utilizing rescaling algorithms \cite{angulo_one_2010} or second order Lagrangian Perturbation Theory in conjunction with time- and scale-dependent transfer functions that provide good approximations to particle trajectories and can be derived from cheap N-body simulations \cite{tassev_estimating_2012}.

Beyond the standard $\Lambda$CDM simulations, there is considerable interest in exploring simulations with alternative gravity laws, such as $f(R)$ theories, that explain cosmic acceleration without DE. In addition to the N-body gravity treatment, such simulations must also solve for the non-linear dynamics of the Chameleon scalar field that screens the gravity law modifications from small scales. In many cases, this is accomplished with adaptive multi-grid relaxation techniques, quite similar to how gravity is treated in AMR hydrodynamics codes. Promising initial progress is already being made in this area \cite{oyaizu_nonlinear_2008,zhao_n-body_2011,brax_systematic_2012,li_nonlinear_2012}. \\

On cluster and galactic scales, efforts will be focused on the substructure problem, and will attack the problem from three directions:

\begin{enumerate}[i)]

\item \underline{Higher resolution}

  Given that the CDM substructure hierarchy extends for 10 - 15 orders of magnitude below current resolution limits, direct numerical simulations will realistically not be able to resolve the full hierarchy even in the intermediate future. Nevertheless, pushing on resolution in zoom-in simulations of individual halos (as in the Via Lactea and Aquarius projects) is a worthwhile goal, for the following reasons:

  \begin{itemize}

  \item Stars in the newly discovered ultra-faint dwarf galaxies of the Milky Way are typically only found out to $\lesssim 100$ pc from the dwarf's center. In order to compare stellar kinematics in these systems with predictions of central densities and profile slopes from CDM simulations, it will be necessary to push to a mass and force resolution of $\sim 100 \Msun$ and $\sim 10$ pc. Note that baryonic physics modifications are expected to be less important in the low mass ($<10^9 \Msun$) host halos of ultra-faints \cite{governato_cuspy_2012}, and so DM-only simulations should still provide valuable predictions.

  \item The DM annihilation boost factors from substructure depend sensitively on the properties of the subhalo population below current simulation's resolution limit (\S\ref{sec:indirect} ix). Since the power spectrum of DM density fluctuations is not truly scale invariant, one might expect quantitative changes in subhalo properties at lower masses. Being able to resolve the abundance, spatial distribution, and density profiles of $<10^5 \Msun$ subhalos would help to clarify the relevance of substructure boost factors.

  \item The phase-space structure in the local neighborhood is only beginning to be resolved by current simulations. Higher resolution simulations will provide a somewhat more fine-grained view, and will additionally resolve more tidal streams from disrupted subhalos. This will allow a better assessment of the importance of velocity substructure for the intepretation of DM direct detection results. A caveat is that baryonic affects are likely to be very important.

  \end{itemize}

The ``Silver River'' simulation is an example of an on-going effort in this class. With a particle mass of $100 \Msun$ and a force softening of 27 pc, this simulation is pushing the frontier in Galactic zoom-in simulations. The run has progressed to $z \approx 5$ and will be completed to $z=0$ in the next 1 - 2 years, pending computational resource support from national supercomputing centers (30 - 40 million core-hours required).

\item \underline{Different host halos}

  The total number of ultra-high resolution ($>10^9$ particles) simulations of individual halos is still quite small: one cluster simulation (Phoenix) and three galactic scale ones (Via Lactea II, Aquarius, and GHalo), and all four of these simulations have been run with only two codes, \textsc{Pkdgrav2} and \textsc{Gadget}. The six lower resolution Aquarius simulations and Millennium-II notwithstanding, the question of cosmic variance, of how much halo-to-halo scatter there is in the substructure population, is not fully settled: see for example \cite{ishiyama_variation_2009}, who find a considerably larger halo-to-halo scatter in the subhalo abundance than what is seen in Aquarius and Millenium-II. Trends with host halo mass and accretion history are equally important and in need of further study. These questions are of particular relevance to the Too Big To Fail problem (\S\ref{sec:CDM_challenges}), where even a factor of a few variation would go a long way towards a resolution \cite{purcell_bailing_2012}. Lastly, current simulations don't properly account for the immediate environment of the Milky Way, because they typically don't have a nearby massive M31-like companion. 

  The Rhapsody Cluster Resimulation Project \cite{wu_rhapsody:_2012} is an early effort along these lines, consisting of 96 zoom-in simulations of cluster scale halos with $\sim 10^7$ particles per halo. Simulation suites of hundreds of Milky Way-like halos with varying masses and accretion histories are also currently being pursued. Capturing more of the Local Group structure can be achieved either through constrained realization simulations \cite{gottloeber_constrained_2010} or by picking suitable halo configurations from an ensemble of lower resolution simulations (S.~Garrison-Kimmel et al., work in progress).

\item \underline{Alternate DM models}

  Going beyond the cold and collisionless DM paradigm, sharpening the predictions that WDM or SIDM models make will be another promising avenue of exploration. In models with a cutoff in the density power spectrum it will be interesting to study how halo formation is suppressed right around the free streaming scale. This is numerically challenging, due to the extremely high resolution required to push the artificial fragmentation to small enough scales. If this challenge can be overcome, it may be feasible to conduct ultra-high resolution Milky Way-scale simulations (like Via Lactea II or Aquarius A-1) with one of the alternative DM models. This would be very useful for comparing with stellar kinematic data in MW dwarf satellite galaxies and with data from future strong lensing surveys of flux ratio anomalies. It may also be of interest to simulate not generic WDM or SIDM, but specific well-motivated particle physics models, for example the Atomic DM model \cite{cyr-racine_cosmology_2012} with its baryonic-like small scale power spectrum oscillations.

\end{enumerate}

\subsection{Simulations Including Baryons Physics}

We anticipate that improvements in the treatment of baryonic physics processes will be a major focus in cosmological simulations over the next decade and beyond. As detailed in \S\ref{sec:DM+hydro}, there are numerous conceptual and technical challenges in properly implementing baryonic physics. Yet these are important problems to tackle, since they have potentially far reaching consequences for our understanding of the distribution of DM in and around galaxies, groups, and clusters, and directly affect expectations for indirect and direct detection experiments.

On cosmic scales, it will be crucial for numerical simulations to quantify how baryonic physics modifies the matter power spectrum \cite{semboloni_quantifying_2011}, how it affects the bias between galaxies and DM halos, and how these effects impact cosmological parameter estimations from upcoming surveys. For individual cluster simulations, it is imperative to establish how reliable X-ray and Sunyaev-Zeldovich measurements of halo masses are \cite{battaglia_cluster_2012}, and how well cluster mass-observable relations can be calibrated. In addition to radiative cooling, star formation, and stellar feedback, such simulations must account for magnetic fields and anisotropic conduction, non-thermal pressure support from turbulence and cosmic rays, as well as AGN feedback.

On galactic scales, one of the most urgent questions that numerical simulation must strive to clarify is under what conditions adiabatic contraction steepens the central DM density profile, and whether this effect can be overcome by feedback processes that may redistribute large amounts of DM and result in shallower or even cored density profiles. If both processes occur in nature, which one dominates, and how does the answer depend on halo mass, environment, and cosmic time? It seems clear from past work that simulations with cooling, but little or inefficient supernova (SN) feedback, exhibit a substantial amount of adiabatic contraction of the DM halo. On the other hand, these simulations typically also suffer from a baryonic over-cooling problem, and produce too many stars, that are too centrally concentrated. Some form of stellar feedback thus appears to be a necessity. Whether the resulting regulation of star formation is accompanied by a removal of substantial amounts of DM from the central regions is an open question.

When efficient SN feedback is imposed ``by hand'', either by artificially turning off gas cooling, or by employing unphysically high star formation or SN energy injection efficiencies, or by hydrodynamically decoupling wind particles from the surrounding gas, substantial redistribution of DM has been observed in a variety of simulations. But how realistically are these ad-hoc implementations capturing the actual physical processes occuring in star forming regions inside giant molecular clouds? Is the resulting DM cusp flattening a robust outcome? The answers await more sophisticated treatments of star formation and feedback processes. Promising directions of future investigations along these lines include, but are not limited to:
\begin{itemize}
\item Suppressing star formation in low mass and low metallicity systems by using H$_2$-regulated star formation prescriptions rather than (or in conjunction with) SN feedback \cite{gnedin_modeling_2009,kuhlen_dwarf_2012};
\item Accounting for radiation pressure from young massive stars \cite{hopkins_self-regulated_2011}, which imparts momentum into the surrounding gas and can increase the efficiency of subsequent SN feedback without resorting to artificial enhancement;
\item Modeling non-thermal support provided by unresolved turbulence, magnetic fields, and cosmic ray propagation \cite{jubelgas_cosmic_2008,scannapieco_simulating_2010}.
\end{itemize}
In general more sophisticated treatments of star formation and feedback physics will require resolving the hydrodynamic component of the simulations with parsec scale resolution (the DM can be treated with coarser resolution for these purposes). This is at least a factor of ten higher than what is currently achievable in cosmological zoom-in simulations. Full-box simulations with this resolution will be limited to the high redshift domain, and thus it will be challenging to obtain large samples of realistically simulated galaxies for comparison with local observations.

It is commonly hoped that it will eventually be possible to include realistic star formation and feedback prescriptions in low resolution, large volume, full-box simulations by calibrating them to the results from much higher resolution zoom-in, or even isolated, simulations, in which a more accurate treatment of the relevant physics is possible. However, it is yet to be demonstrated that this approach will be feasible. Recent algorithmic developments in creating initial conditions (by generating white noise fields hierarchically in real space \cite{hahn_multi-scale_2011,angulo_scaling_2012}) are now enabling simulations of unprecedented dynamic range. The MUSIC code \cite{hahn_multi-scale_2011} has demonstrated high accuracy in reproducing in nested simulations the same detailed structures as in corresponding high uniform resolution full-box runs. This provides the basis for the Cosmic Renaissance (CORE) Project (T.~Abel et al., work in progress), which aims to conduct a large set of self-consistent nested simulations over a wide range of scales, from extreme zoom-ins on the formation of the first stars in a volume the size of the observable Universe to increasingly less focused and more coarsely resolved simulations of the formation of galaxies, clusters, and the large scale structure of the Universe.

Lastly, it is also necessary to understand how these numerical results depend on which hydrodynamic technique and indeed which code was used. The Aquila Code Comparison Project \cite{scannapieco_aquila_2012} was an important step in this direction, but owing to its approach of allowing simultaneous variations in code and feedback physics implementations, it was difficult to disentangle which factor the observed differences can be attributed to. Nevertheless, it clearly demonstrated that the current generation of cosmological hydrodynamic simulations are not yet able to uniquely predict the properties of the baryonic component of a galaxy forming in a fully specified dark matter accretion history. The recently initiated Santa Cruz High-resolution Galaxy Simulation Comparison Project\footnote{https://sites.google.com/site/santacruzcomparisonproject/} is a similar effort, focusing on $\lesssim 100$ pc resolution galaxy formation simulations with a wide range of current state of the art codes representing SPH, AMR, and Moving Mesh techniques.

\subsection{Computational Trends and Algorithmic Advances}

\subsubsection{Processing}\label{sec:processing}

Over the last decade there have been significant advances in High Performance Computing (HPC) available to astrophysical research. The top 20 machines in the June 2012 TOP500 list\footnote{http://www.top500.org/list/2012/06/100} have all exceeded the petaflops barrier, and utilize upwards of 100,000 cores to do so (\textit{Sequoia}, the current No.1 has 1,572,864 cores). Supercomputers are expected to reach the exaflop scale during the next decade. This will enable us to carry out much more detailed modeling of the problems described in this manuscript, but it also poses serious challenges for the development of numerical codes and algorithms capable of fully exploiting new technologies.

\begin{enumerate}[i)]
\item \underline{Scalability and performance}

One characteristic of current and future supercomputers is the very large number of computing cores. The most obvious problem concerns how to distribute the computational load over an increasingly large computational domain, while at the same time keeping communication time to a minimum. Preserving adequate balance in the CPU and load requirements will require development and improvements in all algorithms present in N-body calculations, in order to achieve scalability to millions of compute tasks. 

Another interesting aspect of supercomputing in the next decade will be the architecture of their Central Processing Units (CPU). Modern compute nodes contain multiple processing cores, each of which have their own sets of instructions and cache, but have access to common memory (RAM). In the near future we can expect nodes with hundreds of such cores, a feature that needs to be exploited by simulation codes via mixed parallelization schemes, in which the commonly used Message Passing Interface (MPI) model for distributed memory is combined with shared memory parallelism, via e.g. OpenMP or Posix threads. Some existing N-body solvers already take advantage of SIMD (Single Instruction Multiple Data), which allows parallel vector calculations in the CPU. Speed-ups can be factors of a few, however exploiting SIMD requires machine-dependent code.

In recent years, Graphical Processing Units (GPU) have become a competitive alternative to CPUs for intensive numerical tasks. Initially developed for rapid and highly parallelized manipulation of computer graphics, they have also found wide use for general-purpose computation. For certain algorithms able to take advantage of data-parallelism, GPUs can result in very significant speed-ups of factors of ten or more. HPC is already beginning to embrace this trend, with 3 of the top 10 supercomputers (\textit{Tianhe-1A}, \textit{Jaguar}, and \textit{Nebulae}) taking advantage of GPUs. A downside of using GPUs is the comparatively small amount of memory available on board, requiring a lot of slow data transfer from CPU to GPU and back. Nevertheless, there have been efforts to implement widely used algorithms to take advantage of GPUs, for instance Fast Fourier Transforms, kd-Trees, as well as visualization and analysis tools. Custom code modifications can be performed using extensions to the C programming language, like CUDA (specific to NVIDIA GPUs) and OpenCL (multiple architectures). As graphics cards improve in reliability and performance, cosmological N-body codes will likely take advantage of this technology (see e.g. \cite{hamada_42_2009,jetley_scaling_2010,santanu_optimising_2012}), most likely in programming models that employ hybrid algorithms for both CPUs and GPUs.

Perhaps a point of convergence between CPUs and GPUs will be in hardware such as the Intel MIC cards (Many Integrated Core). This product, resulting from Intel's Larrabee research project, is essentially a GPU-CPU hybrid and functions as a co-processor for HPC. It is expected to become available by the end of 2012 and could become an alternative to GPUs with a simpler programming model.

\item \underline{Big Data}

As we have seen, state of the art simulations already generate petabytes of data. Handling
this data is becoming an increasingly difficult problem and we can only
anticipate that it will get worse during the next decade. One limiting aspect is
the disk I/O speed; even with parallel distributed filesystems (e.g. Lustre) it does not scale with the
size of the supercomputer, and thus disk I/O can take a considerable fraction
of the runtime of a simulation. Another aspect is how to manage and store 
extremely large datasets, especially with collaborations commonly spread across the world and the considerable cost of dedicated storage hardware. Simply migrating the data from large supercomputing centers to smaller local facilities can be difficult and time consuming.

These are serious challenges for computational cosmologists, and strategies will need to be developed to cope with these issues. One natural development already in practice is to merge analysis and visualization software with the simulation code so as to reduce as much as possible the output of data during runtime and the handling afterwards.\footnote{This is also one of the stated development goals of the yt Project \cite{turk_high-performance_2011}, http://yt-project.org/.} For large DM simulations this may mean identifying and extracting (sub)halo information, constructing merger trees and lightcones on-the-fly, and storing only the resulting reduced data. However, it is never possible to anticipate all the uses of a simulation, and this motivates saving at least some full outputs. Novel data compression techniques are a promising way forward. The idea is that data analysis often does not require the same precision as run-time computation. Substantial compression of the output can the be achieved either by neglecting irrelevant bytes or by spatial averaging (a compression approach widely used in image and video displays). For instance, the data can be stored in a hierarchical tree fashion, where the initial bytes provide a coarse representation of the dataset, and subsequent bytes provide refinement in specific quantities and/or spatial regions. These would alleviate access and reading time for postprocessing, at a cost of giving up the ability to restart simulations and with file formats very specific to a given simulation.

The distribution of data worlwide is a serious but important problem,
as serving data products is necessary and desirable for scientific 
progress. In the last few years, an increasingly popular option has been the developments of internet databases providing reduced data products, for example from the Millennium and MultiDark simulation databases\footnote{http://www.mpa-garching.mpg.de/millennium/ and http://www.multidark.org/MultiDark/}. Another alternative is to store the data itself in a distributed fashion on the internet in the ''cloud'' and using Hadoop for distributed data analysis.

\end{enumerate}

\subsubsection{Novel Approaches}

\begin{enumerate}[i)]

\item \underline{New Gravity Solvers}

Despite well-known problems with the N-body treatment of cosmological structure formation (e.g. two body relaxation and artificial fragmentation), there has been little progress in how to solve the collisionless Boltzmann equation with self-gravity (Poisson-Vlasov equations). As the accuracy demanded of cosmological simulations increases, it is likely that the short-comings of current N-body schemes will eventually become limiting. This motivates improvements in existing treatments, such as the introduction of adaptive gravitational softening \cite{price_energy-conserving_2007,iannuzzi_adaptive_2011} or improved density estimators based on following the distortion of a tetrahedral tessellation of the initial density field \cite{shandarin_cosmic_2012,abel_tracing_2011}, as well as radical departures from the Monte Carlo N-body approach, such as full six-dimensional Vlasov solvers \cite{yoshikawa_direct_2012} or employing kinetic theory \cite{ma_cosmological_2004}.

\item \underline{New Programming Languages - the need for parallelism}

Parallelism and HPC become increasingly important also for industry and
business applications, where server farms and computing cluster with tens of
thousands of cores are now used for large-scale data mining and reduction
problems. Furthermore, since the break-down of Moore's law for the raw core
processor performance, developers in all fields, ranging from mobile devices to
high performance cluster, need to start thinking parallel. It is therefore not
surprising that new programming languages, paradigms, and models are now
emerging, which have parallelism more naturally built in. 

One example is Unified Parallel C (UPC) as an extension of the ISO C 99
programming language designed for high-performance computing on large-scale
parallel systems. This includes architectures with common global address spaces
(SMP and NUMA) and also distributed memory systems, which are more common for
cosmological applications. The main idea of UPC is to present the developer
with a single shared, partioned address space, and a Single Instruction
Multiple Data (SIMD) programming model. UPC is a specific implementation of the
more general class of PGAS (partitioned global address space) programming
languages.  Other languages which follow the PGAS principles are Co-array
Fortran, Titanium, Fortress, Chapel, X10 and Global Arrays.  PGAS is based on
two main concepts:  multiple execution contexts with separate address spaces
and access of memory locations on one execution instance by other execution
instances. It is very likely, that in the near future simulation codes will
start using some of these language. 

Besides the actual simulation software, data processing software also becomes
increasingly complex and requires efficient parallelization.  Large parts of
available post-processing pipelines are programmed in general-purpose,
interpreted high-level programming languages like Python, which is slowly
replacing older IDL implementations. Both languages were essentially designed
without a particular focus on parallelism, though it is possible to use them
for parallel processing. But new languages are also arising in this field,
which will likely replace Python or IDL as the de-facto standards in the near
future. For example, Julia is new a high-level, high-performance dynamic
programming language for technical computing, which provides a sophisticated
compiler, distributed parallel execution, and an extensive mathematical
function library. Altough it is a LLVM-based just-in-time (JIT) compiler based
language it reaches nearly C/C++ performance due to heavy optimization
outperforming IDL, MATLAB, R, Python, Octave by a large factor. Most
importently it has parallelism naturally build in, which makes it suitable for
large-scale data mining and reduction on huge compute clusters. 

UPC, as a PGAS parallel programming language, and Julia, as a high-performance
dynamic programming language, are just two examples and it is very likely that
the programming language landscape will move significantly more towards HPC and
parallel computing in the very near future mainly driven by industry needs.

\end{enumerate}

\section{Conclusions}

Two of the greatest mysteries of contemporary physics are the nature of dark matter and of dark energy. As we have seen over and again in this work, experimental and observational efforts to get at the answers to these questions are intimately tied to predictions from cosmological DM simulations. The simulation field, however, is rapidly evolving, as computational resources continue to grow, enabling larger, more complicated, and more realistic simulations to be performed. As a result, it has become more difficult for physicists not directly involved with simulations to follow the progress of the field, and to stay up to date with the evolving implications for studies of DM and DE. Motivated by this realization, we have in this review attempted to provide an overview of the current state of the field of cosmological DM simulations, with a particular emphasis on the connection to DM detection experiments and observational probes of DE. We have summarized the successes and accomplishments of the current generation of simulations, but also detailed their problems and short-comings, as well as the challenges faced by the next generation of simulations in the coming decade.

For the DE problem, the greatest need appears to be a tremendous increase in the number of very high resolution, very large volume, DM-only simulations, scanning over cosmological parameters in order to allow brute-force Monte-Carlo estimates of the parameter errors and their co-variances for future DE surveys. At the same time, cosmic scale simulations will continue to investigate the effects of baryonic physics, non-Gaussian initial conditions, and modified laws of gravity.

For the DM problem, on the other hand, it seems that the DM-only approach by itself is nearing the end of its usefulness -- baryonic physics effects are simply too important to neglect on galactic scales. There is some room for improvement in the DM-only approach, in particular for the internal structure of subhalos, for exploring cosmic variance and host halo mass dependency, and for departures from the cold and collisionless DM assumption. However, the main driver of the field must be studies of baryonic effects on the distribution of DM in halos, since these have the potential to profoundly alter expectations for DM detection.

The next decade promises to be filled with exciting challenges and the potential for great discoveries. It is safe to say that the field of Numerical Simulations of the Dark Universe will not run out of things to do.

\bigskip

\section*{Acknowledgments}

We are very grateful to Tom Abel, Jean-Michel Alimi, Michael Boylan-Kolchin, Francis-Yan Cyr-Racine, Liang Gao, Phil Hopkins, Anatoly Klypin, Chung-Pei Ma, Kristin Riebe, Graziano Rossi, Uros Seljak, Kris Sigurdson, Joachim Stadel, Romain Teyssier, and Martin White for valuable discussions and assistance with the simulation survey. This work has made extensive use of NASA's Astrophysics Data System.

\bibliographystyle{elsarticle-num}
\bibliography{FoNS}

\end{document}